\documentclass[journal=aamick,manuscript=article]{achemso}

\usepackage[version=3]{mhchem} 

\usepackage[normalem]{ulem}
\usepackage{cancel}
\usepackage{amsthm}

\usepackage[export]{adjustbox}
\usepackage{xcolor}
\definecolor{C0}{HTML}{1F77B4}
\definecolor{C1}{HTML}{FF7F0E}
\definecolor{C2}{HTML}{2ca02c}
\definecolor{C3}{HTML}{d62728}
\definecolor{C4}{HTML}{9467bd}
\definecolor{C5}{HTML}{8c564b}

\def\usetodonotes{} 
\ifdefined\usetodonotes
\usepackage{todonotes}
\usepackage{changes}

\newcommand{\nikhil}[1]{\todo[size=\scriptsize,color=blue!10,bordercolor=blue,linecolor=blue]{NA: #1}}

\newcommand{\tusher}[1]{\todo[size=\scriptsize,color=magenta!10,bordercolor=magenta,linecolor=magenta]{TA: #1}}

\fi
\usepackage{tabularx}
\usepackage{upgreek}
\usepackage{tikz}
\usepackage{soul}
\usepackage{tcolorbox}
\usepackage[scr=esstix,cal=boondox]{mathalfa}
\usepackage{booktabs}
\usepackage{graphicx}
\usepackage{dcolumn}
\usepackage{bm}

\usepackage{siunitx}
\usepackage[skip=0pt]{caption}
\usepackage{subfig}
\usepackage{float}
\usepackage{amsmath}
\usepackage{amsmath}
\usepackage{amsfonts}
\newcommand{\angstrom}{\textup{\AA}}

\setcitestyle{square, numbers, comma, sort&compress, super}
\usepackage{algpseudocode}
\usepackage{algorithm}
\usepackage{latexsym}
\DeclareMathOperator{\divr}{Div}

\usepackage[colorlinks=true]{hyperref}
\hypersetup{ 
    colorlinks=true,       
    linkcolor=red,          
    citecolor=blue,        
    filecolor=magenta,      
    urlcolor=cyan           
}
\usepackage{cleveref}

\newcommand{\psubref}[1]{\protect\subref{#1}}

\newcommand{\Xt}{\bm X_{\rm t}}
\newcommand{\Xb}{\bm X_{\rm b}}

\newcommand{\Ht}{\bm H_{\rm t}}
\newcommand{\Hb}{\bm H_{\rm b}}
\newcommand{\Kt}{\bm K_{\rm t}}
\newcommand{\Kb}{\bm K_{\rm b}}

\newcommand{\phit}{\bm{\phi}_{\rm t}}
\newcommand{\phib}{\bm{\phi}_{\rm b}}
\newcommand{\rt}{\bm r_{\rm t}}
\newcommand{\rb}{\bm r_{\rm b}}

\newcommand{\sref}[1]{Section~\ref{#1}}
\newcommand{\fref}[1]{Fig.~\ref{#1}}
\newcommand{\frefs}[2]{Figs.~\ref{#1} and \ref{#2}}
\newcommand{\divrt}{\divr}
\newcommand{\divrb}{\divr}
\author{Md Tusher Ahmed}
\affiliation
{Department of Mechanical Science and Engineering,
 University of Illinois at Urbana-Champaign, Urbana, IL, USA}
\author{Moon-ki Choi}
\affiliation
{Material Research Laboratory,
 University of Illinois at Urbana-Champaign, Urbana, IL, USA}
\author{Harley T Johnson}
\affiliation
{Department of Mechanical Science and Engineering,
 University of Illinois at Urbana-Champaign, Urbana, IL, USA}
 \alsoaffiliation{
Department of Materials Science and Engineering,
 University of Illinois at Urbana-Champaign, Urbana, IL, USA
}
\author{Nikhil Chandra Admal}
\affiliation
{Department of Mechanical Science and Engineering,
 University of Illinois at Urbana-Champaign, Urbana, IL, USA}
\email{admal@illinois.edu}

\title[An \textsf{achemso} demo]
  {Quantifying superlubricity of bilayer graphene from the mobility of interface dislocations}

\begin{document}

\begin{abstract}
  Van der Waals (vdW) heterostructures subjected to interlayer twists or heterostrains demonstrate structural superlubricity, leading to their potential use as superlubricants in micro- and nano-electro-mechanical devices.  However, quantifying superlubricity across the vast four-dimensional heterodeformation space using experiments or atomic-scale simulations is a challenging task. In this work, we develop an atomically informed dynamic Frenkel--Kontorova (DFK) model for predicting the interface friction drag coefficient of an arbitrarily heterodeformed bilayer graphene (BG) system.  The model is motivated by MD simulations of friction in heterodeformed BG. In particular, we note that interface dislocations formed during structural relaxation translate in unison when a heterodeformed BG is subjected to shear traction, leading us to the hypothesis that the kinetic properties of interface dislocations determine the friction drag coefficient of the interface.
The constitutive law of the DFK model comprises the generalized stacking fault energy of the AB stacking, a scalar displacement drag coefficient, and the elastic properties of graphene, which are all obtained from atomistic simulations. 
Simulations of the DFK model confirm our hypothesis since a single choice of the displacement drag coefficient, fit to the kinetic property of an individual dislocation in an atomistic simulation, predicts interface friction in any heterodeformed BG. By bridging the gap between dislocation kinetics at the microscale to interface friction at the macroscale, the DFK model enables a high-throughput investigation of strain-engineered vdW heterostructures.
\end{abstract}
\section{Introduction}
In the era of micro- and nano-electro-mechanical devices, reducing friction and wear at the interfaces of miniature devices has become a prominent research topic. Lubricating interfaces in micro and nanoscopic devices using conventional lubricants is often challenging, creating the need for a new class of superlubricants that yield nearly zero interface friction \citep{williams2006tribology,kim2007nanotribology,hod2018structural,briscoe2006boundary}. Suberlubricants are renowned for showing enhanced resistivity to thermal and chemical degradation as well as superior wear resistance due to their atomically thin geometry \citep{berman2014graphene,baykara2018emerging}. Bilayer 2D materials are some of the most promising candidates for superlubricants widely used in nanoelectromechanical systems \citep{vazirisereshk2019origin,song2018robust,leven2013robust,li2017superlubricity,buch2018superlubricity,kawai2016superlubricity}. Bilayer graphene (BG), in particular, is an exceptional superlubricant owing to its low friction and excellent mechanical, thermal, and electrical properties \citep{baykara2018emerging}. Furthermore, the friction coefficient of a BG can be adjusted by altering twist angles and strain states \citep{dienwiebel2005model, androulidakis2020tunable}, enhancing its versatility for diverse applications.

Following the investigation of superlubricity by 
\citet{dienwiebel2004superlubricity} in rotationally misaligned graphite layers, numerous studies on misaligned homo- and hetero-structures \citep{ruan2021robust,ferrari2021dissipation,wei2022interlayer} have recognized the twist angle and heterostrain\footnote{In a heterostrained configuration, one of the layers is strained relative to the other. Twist angle and heterostrain are usually prescribed relative to the lowest energy stacking.} dependence of friction. Adopting our terminology in \citep{ahmed2024bicrystallography}, we will use the umbrella term \emph{heterodeformations} to refer to interlayer twists and heterostrains. Depending on the magnitude of the relative twist and/or heterostrain, there can be a substantial variation in friction forces giving rise to friction anisotropy \citep{weick2001anisotropic}.   

However, experimentally measuring interface friction across numerous heterodeformations is highly challenging due to the enormity of the four-dimensional heterodeformation space.\footnote{If the two lattices of a heterostructure are subjected to uniform deformation gradients $\bm F_1$ and $\bm F_2$, then $\bm F_1^{-1}\bm F_2$ describes the relative deformation and its representation as a $2\times 2$ matrix accounts for the four dimensions.} In this regard, molecular dynamics (MD) simulations provide a viable alternative. However, MD simulations are computationally expensive because --- a) the time scale is restricted to a few microseconds, often limiting the study to sliding velocities that are six orders of magnitude larger than those in experiments \citep{wang2024colloquium}; b) periodic boundary conditions (PBCs) enforced to remove edge effects result in large simulations domains for many heterodeformations. The limitations of MD simulations motivate us to develop a high-throughput multiscale framework to characterize friction anisotropy.

The classical Prandtl--Tomlinson (PT) model \citep{prandtl1928gedankenmodell, tomlinson1929cvi, popov2012prandtl} is a continuum model to describe atomic-scale friction in crystalline materials. For instance, the PT model can describe the experimentally observed stick-slip characteristic of friction in bilayer 2D interfaces \citep{dienwiebel2005model, song2018robust}. The PT model's simplicity and adaptability make it a valuable tool for predicting interfacial friction in bilayer 2D materials \citep{zworner1998velocity, andersson2020understanding, song2022velocity, huang2024general}. A key feature of the PT model is the sinusoidal potential energy representing the interaction between two sliding surfaces. However, since the periodic potential energy depends on the heterodeformation, a single PT model cannot be used to predict friction across multiple heterodeformed bilayers.

The goal of this paper is to develop an atomistically informed dynamic Frenkel--Kontorova (DFK) continuum model of a bilayer to predict interface friction for \emph{any} heterodeformation of the interface. Our model is motivated by the critical role interface dislocations --- observed in heterodeformed bilayers of 2D materials as a consequence of structural relaxation \citep{harley_disloc, ADMAL2022, ahmed2024bicrystallography} --- play while the two layers of a BG slide relative to each other. We show that the DFK model fitted to the kinetic properties of a single dislocation can predict interface friction for arbitrary heterodeformations. We note that the current strategy of predicting friction from the properties of interface dislocations can also be found in the work of \citet{merkle2007predictive} on twist grain boundaries.

The paper is organized as follows. \sref{sec:friction} begins with a review of structural reconstruction and the formation of interface dislocations using MD simulations of BGs subjected to heterodeformations. Subsequently, MD simulations of sliding friction are used to identify the critical role of interface dislocations and their mobility in determining friction.
In \sref{sec:cont_fram}, we present the DFK model for predicting friction in heterodeformed BG. In \sref{sec:det_mobility}, we obtain the kinetic parameters of the DFK model from the kinetics of a single dislocation simulated using atomistics. Subsequently, the fitted DFK model is used to predict interface friction for numerous heterodeformations. 
We summarize and conclude in \sref{sec:conclusion}. 
\\

\noindent
\emph{Notation}:
Lowercase bold letters are used to denote vectors while uppercase bold letters are used to denote second-order
tensors, unless stated otherwise. The gradient and divergence operators are denoted by the symbols $\nabla$ and $\divr$ respectively. We use the symbol $\cdot$ to denote the inner product of two vectors or tensors.
\section{Atomic scale study of friction in heterodeformed BG}
\label{sec:friction}
We use atomistic simulations to investigate friction in a BG subjected to heterodeformations. Simulations are performed using Large-Scale Atomic/Molecular Massively Parallel Simulator (LAMMPS) \citep{thompson2022lammps} and visualized using OVITO \citep{stukowski2009visualization}. We study two heterodeformations -- a) a small twist of $0.760443^\circ$ and b) a small equi-biaxial heterostrain of $0.4219\%$, relative to the AB stacked BG. The two stated heterodeformations belong to a large collection of heterodeformations --- identified by Smith normal form (SNF) bicrystallography \citep{ADMAL2022}\footnote{SNF bicrystallography is a robust framework that leverages integer matrix algebra to automate the generation of rational grain boundaries (GBs) and identify disconnection modes within them. This approach leads to dimension-independent algorithms that are applicable to any crystal system. The SNF bicrystallography framework is implemented as an open-source C++ library, called open Interface Lab (\texttt{oiLAB}), accessible at \url{https://github.com/oilab-project/oILAB.git}.} --- that are amenable to periodic boundary conditions.

\begin{figure}[t]
    \centering
    \subfloat[]
    {
        \includegraphics[width=0.3\textwidth]{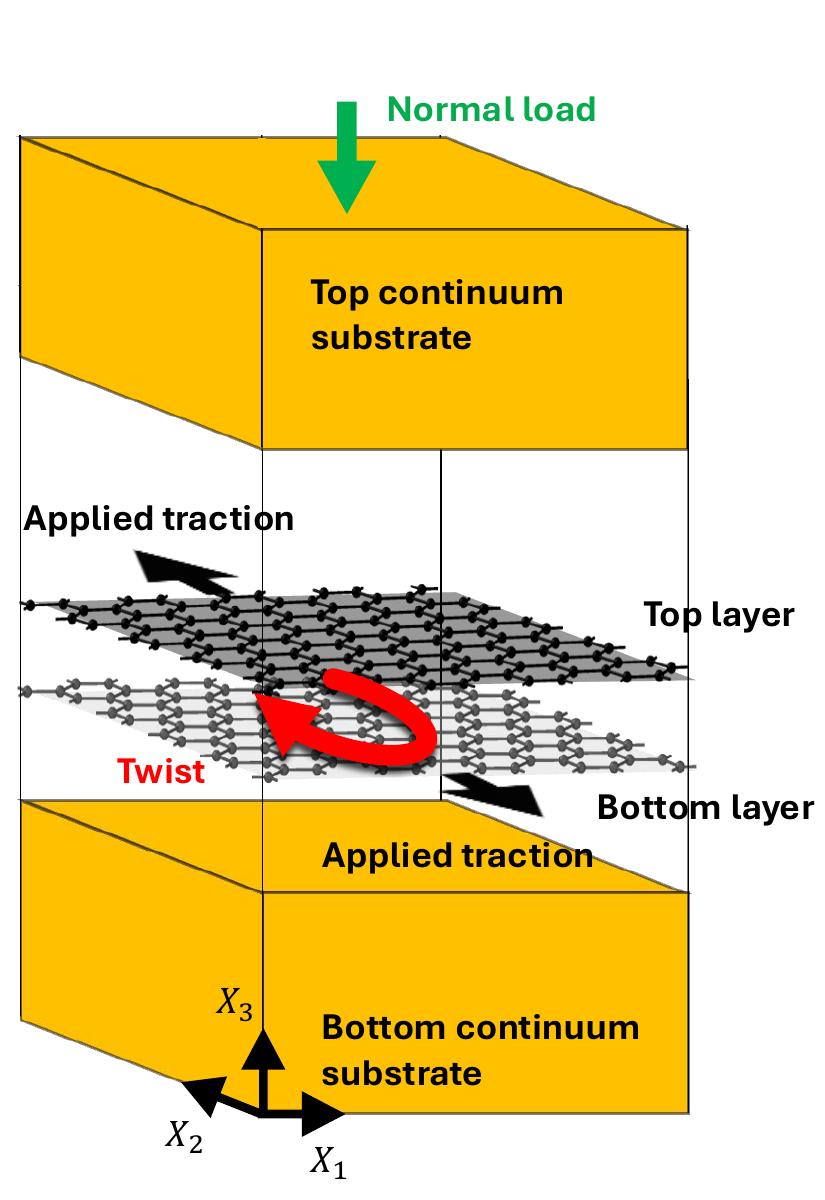}
    \label{fig:schematic_twist}
    }
    \subfloat[]
    {
        \includegraphics[width=0.3\textwidth]{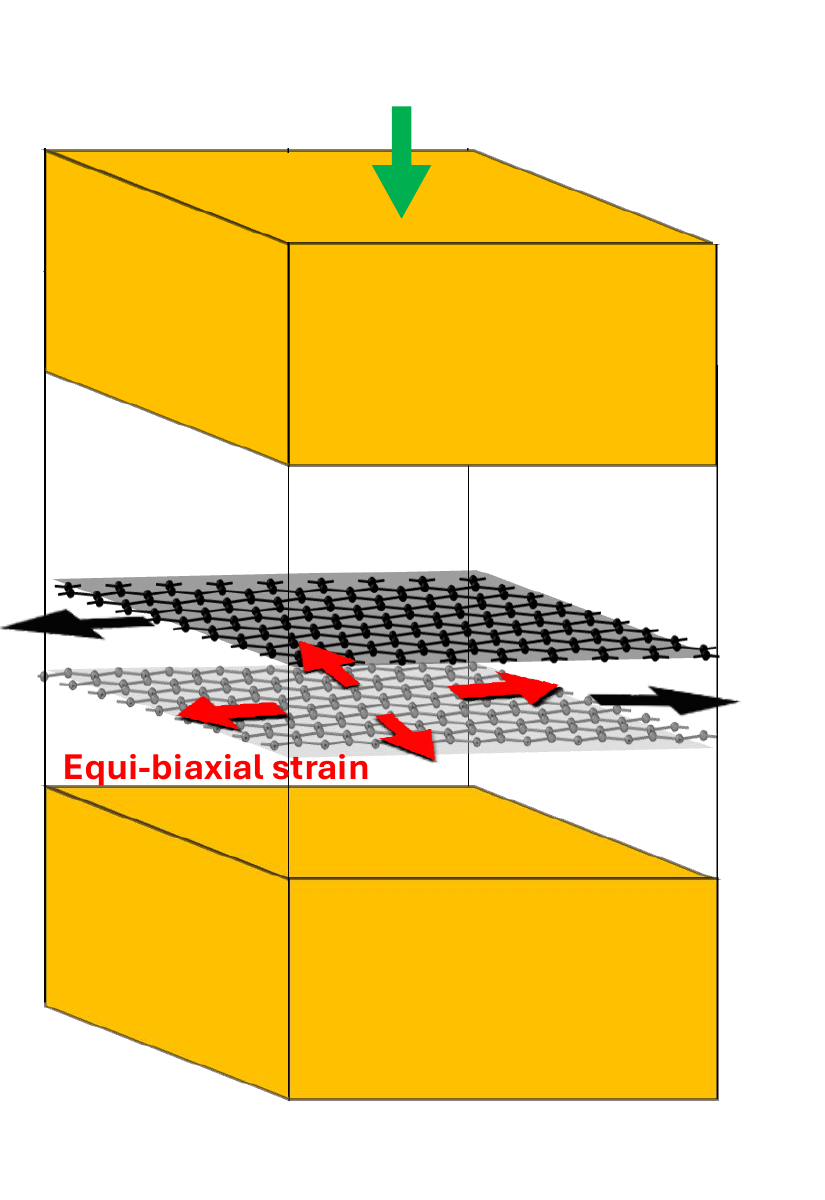}
    \label{fig:schematic_strain}
    }
    \caption{A schematic of atomistic systems used to measure friction in \protect\subref{fig:schematic_twist} twisted BG and \protect\subref{fig:schematic_strain} equi-biaxial heterostrained BG. Continuum substrates, shown in orange, support the bilayer. The black and green arrows represent the applied shear traction forces and normal loads, respectively, while the red arrows indicate the nature of the heterodeformation.}
\label{fig:schematic_cont_sub}
    \end{figure}

The top graphene layer, represented by lattice $\mathcal A$, is constructed using the structure matrix:
\begin{equation*}
    \bm A = \frac{a}{2} \begin{bmatrix}
    0 &  -\sqrt{3}\\
    2 & -1
    \label{eq:A_matrix}
    \end{bmatrix},
\end{equation*}
where the columns of $\bm A$ represent the basis vectors, and $a=2.46~\angstrom$ is the lattice constant of strain-free graphene. The two basis atoms are positioned at coordinates $\left(0,0\right)$ and $\left(\frac{1}{3},\frac{2}{3}\right)$ relative to the basis vectors of $\mathcal A$. The bottom layer, represented by lattice $\mathcal B$, is constructed using the structure matrix $\bm B=\bm F \bm A$. The bilayer is in an AB stacking if $\bm F$ is a rotation of $60^\circ$, and therefore, for the small twist case $\bm F=\bm R(60.760443^\circ)$ and for the small equi-biaxial heterostrain case $\bm F=\bm R(60^\circ) \bm U$, where 
\begin{equation}
    \bm U = 
    \begin{bmatrix}
        1.004219 &  0\\
        0        &  1.004219
    \end{bmatrix}.
    \label{eqn:U_smallTwist}
\end{equation}
In addition to heterodeformations, SNF also yields the following simulation domain vectors 
\begin{subequations}
    \begin{align}
        \text{twist: } \bm b_1 = 185.351 \, \bm e_1, 
        &\quad 
        \bm b_2=92.675 \,  \bm e_1 +160.519\, \bm e_2, \text{ and }\\
        \text{equi-biaxial heterostrain: } \bm b_1 = -585.481\, \bm e_1, 
        &\quad 
        \bm b_2=292.740\,  \bm e_1 - 507.041\, \bm e_2,
    \end{align}
    \label{eqn:pbc_small_strain}%
\end{subequations}
where $\bm e_i$s denote the unit vectors parallel to the global axes $X_1$, $X_2$, $X_3$. Since 2D materials are often supported by 3D substrates \citep{song2018robust,kazmierczak2021strain}, we introduce two continuum substrates that interact with the graphene layers as shown in \fref{fig:schematic_cont_sub}. A continuum substrate serves as a more realistic boundary condition compared to a free-standing or out-of-plane displacement-constrained boundary condition. 

The reactive empirical bond order (REBO) potential \citep{Rebo_Brenner_2002} is used to model the intralayer bonding in each graphene layer while the registry-dependent Kolmogorov--Crespi (KC) potential \citep{ouyang2018nanoserpents} describes the interlayer van der Waals (vdW) interaction. Table~\ref{table:parameters} lists the parameters for the KC potential.
\begin{table}[H]
\centering
\small
\begin{tabular}{c c c c c c c c}
       $C [\si{\meV}]$ & $C_0 [\si{\meV}]$       & $C_2 [\si{\meV}]$ &     $C_4 [\si{\meV}]$  &   $A [\si{\meV}]$  & $\delta [\si{\angstrom}]$  &   $\lambda [\si{\per \angstrom}]$ &   $z_0 [\si{\angstrom}]$ \\
\hline
$\num{6.678908e-4}$ & $\num{21.847167}$ & $\num{12.060173}$ & $\num{4.711099}$ & $\num{12.660270}$ & $\num{0.771810}$ &  $\num{3.143921}$ & $\num{3.328819}$.
\end{tabular}
\caption{Parameters of the KC potential}
\label{table:parameters}
\end{table}
    
\noindent The graphene layers interact with the continuum substrates through the Lennard-Jones potential \citep{lennard1924determination}. Section S$-1$ of the supporting information describes The implementation of the graphene-substrate interaction is described in Section S$-1$ of supporting information.

Our simulation proceeds in two stages. In the first stage, the system is allowed to relax using the fast inertial relaxation engine (FIRE) algorithm \citep{bitzek2006fire} with an energy tolerance of $\SI{1e-20}{\eV}$ and a force tolerance of $\SI{1e-20}{\eV \per \angstrom}$. In the second stage, we measure friction in an NVT ensemble by introducing a relative velocity between the two graphene layers by subjecting the system to in-plane shear traction and an out-of-plane pressure of $2.57$ MPa similar to the experimental study of \citet{qu2020origin} at a low temperature ($T=0.01~K$). The shear traction is enforced by applying equal force to every atom of the top layer in a $X_i$ ($i=1$ or 2) direction while providing the same force to the atoms in the bottom layer in the negative $X_i$ direction. The out-of-plane pressure is enforced by applying equal force to every atom of the top layer in the negative $X_3$ direction.

\subsection{Interface dislocations-mediated atomic reconstruction in heterodeformed BG}
\label{sec:atomic_recons}
It is well known that AB-stacked BG, when subjected to a small heterodeformation, undergoes structural relaxation mediated by interface dislocations \citep{harley_disloc,Annevelink_2020,ahmed2024bicrystallography}. In this section, we report how twist and equi-biaxial heterodeformations result in a markedly distinct arrangement of interface dislocations during energy minimization.

\begin{figure}
    \centering
    \includegraphics[width=0.6\linewidth]{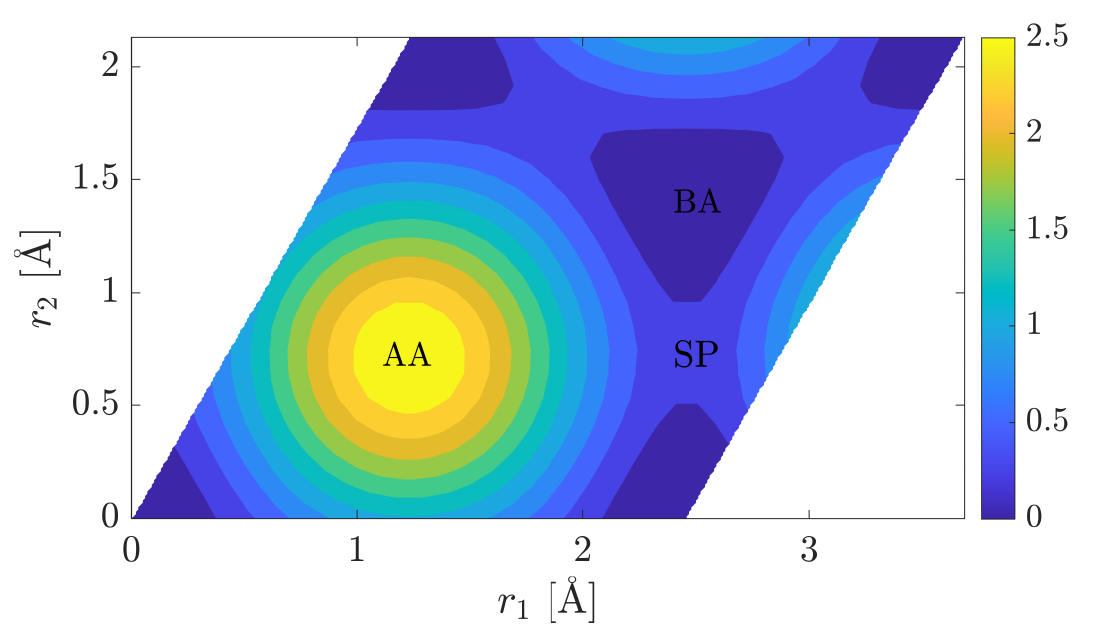}
    \caption{Generalized stacking fault energy [in meV$\si{\angstrom}^{-2}$] plot of AB stacked BG where, $(0,0)$ represents the AB stacking.}
    \label{fig:GSFE_atom}
\end{figure}

The interfacial energy density of the AB stacking corresponds to the energy well at the origin in the generalized stacking fault energy plot (GSFE), shown in \fref{fig:GSFE_atom}. Atomic relaxation following a uniform heterodeformation stems from the spatially varying relative translation of the two layers. The bilayer responds to minimize the interface energy by relaxing to the nearest low-energy stacking, resulting in AB and BA stacked regions separated by interface dislocations. Evidently, the displacement jumps across a dislocation, which constitutes its Burgers vector, is the vector connecting the two adjacent AB and BA regions in the GSFE, whose magnitude is $\SI{1.42}{\angstrom}$. Therefore, the magnitude of the Burgers vector is always equal to $\SI{1.42}{\angstrom}$ for small heterodeformations relative to the AB stacking and such defects are classified as partial dislocations.\footnote{In a previous work \citep{ahmed2024bicrystallography}, we showed that structural relaxation also occurs under small heterodeformations relative to the $\SI{21.786789}{\degree}$ twisted configuration. Interestingly, in such cases, the Burgers vector magnitude is not equal to $\SI{1.42}{\angstrom}$.}

\begin{figure}[t]
    \centering
    \subfloat[]
    {
        \includegraphics[width=0.55\textwidth]{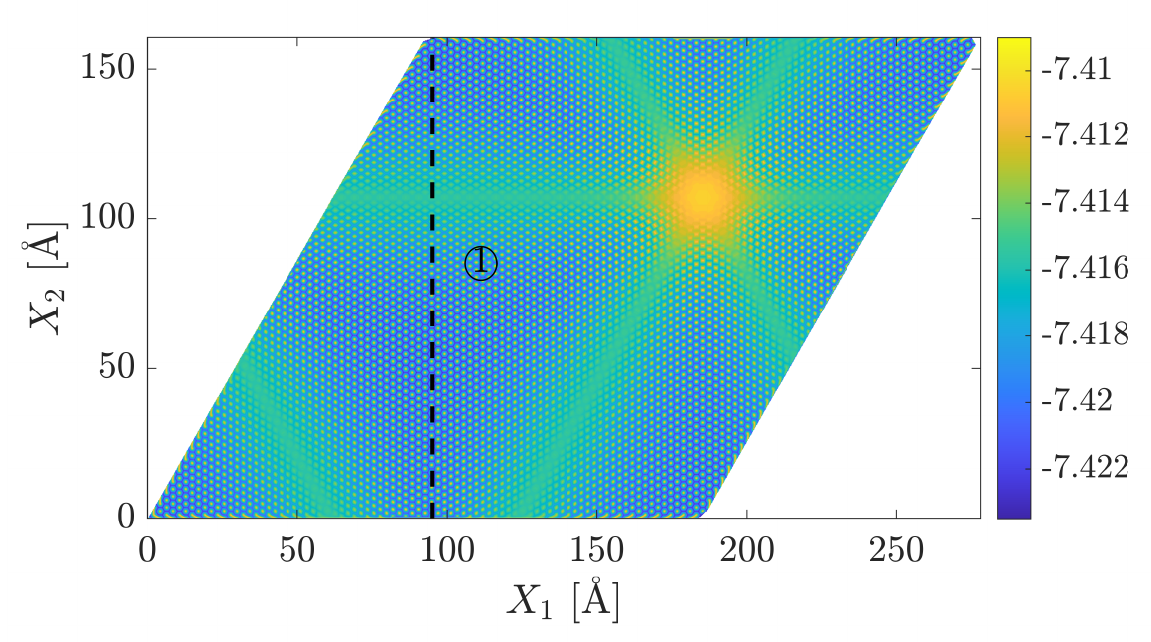}
        \label{fig:relax_twist_en}
    }
    \subfloat[]
    {
        \includegraphics[width=0.45\textwidth]{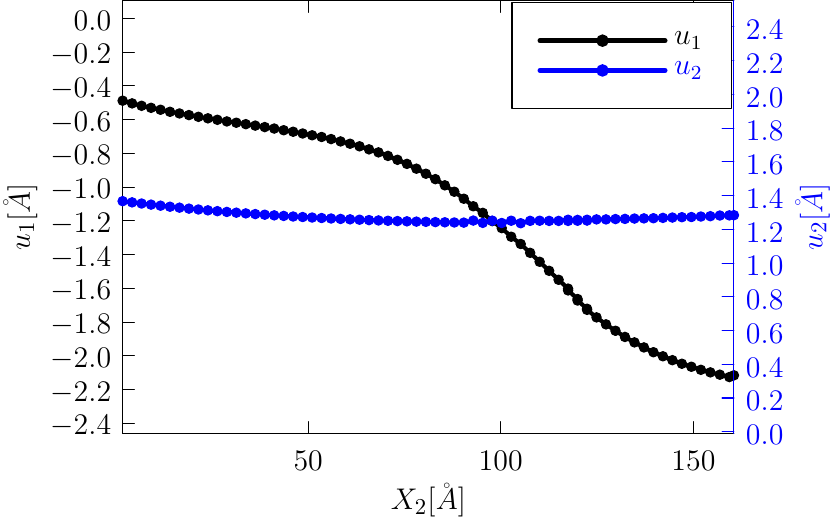}
        \label{fig:relax_twist_scan}
    }
    \caption{Atomic reconstruction in a $\SI{0.76}{\degree}$ twisted BG. \protect\subref{fig:relax_twist_en} Atomic energy per atom [in $\si{\eV}$] plot showing a triangular network of interface dislocations. \protect\subref{fig:relax_twist_scan} Displacement components $u_1$ and $u_2$ relative to the AB stacking [in $\si{\angstrom}$], measured along the line \textcircled{1} shown in \protect\psubref{fig:relax_twist_en}, signify the screw characteristic of dislocations in a twisted BG.}
\label{fig:twist_disloc_character}
    \end{figure}

\fref{fig:relax_twist_en} shows a plot of atomic energy per atom in a post-relaxed twisted BG. Dislocation lines with considerable elastic and interfacial energies separate the low-energy AB and BA triangular domains. In \fref{fig:relax_twist_scan}, we measure the displacement jump across the dislocation line \textcircled{1} to infer the edge/screw characteristic of the dislocation. The variable $u_i$ denotes the displacement component along $X_i$ direction measured relative to the AB stacking. Since the jump in $u_2$ across the dislocation is negligible, it follows that the displacement jump is parallel to the dislocation, which implies the dislocations in a twisted BG have a screw character. 

\begin{figure}[h!]
    \centering
        \subfloat[]
    {
        \includegraphics[height=0.30\textwidth]{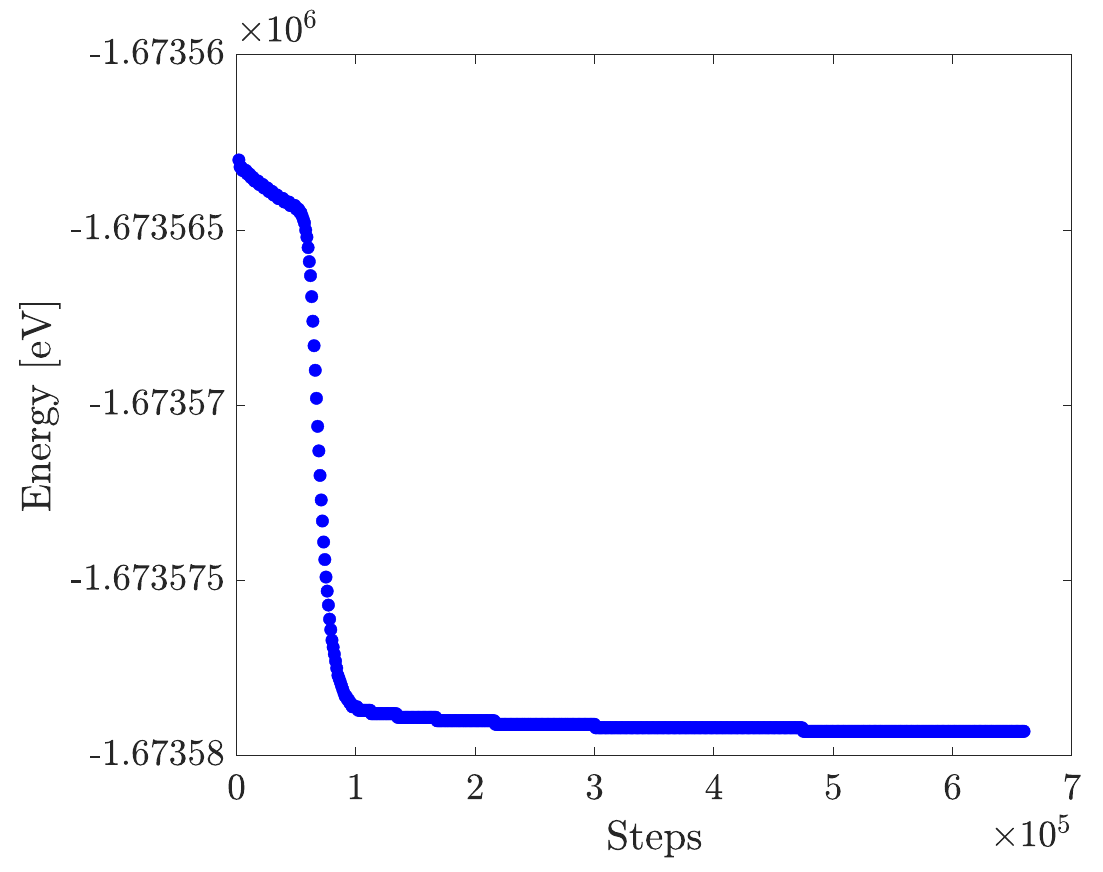}
        \label{fig:energy_landscape}
    }
    \\
    \subfloat[configuration at step $10000$]
    {
        \includegraphics[height=0.25\textwidth]{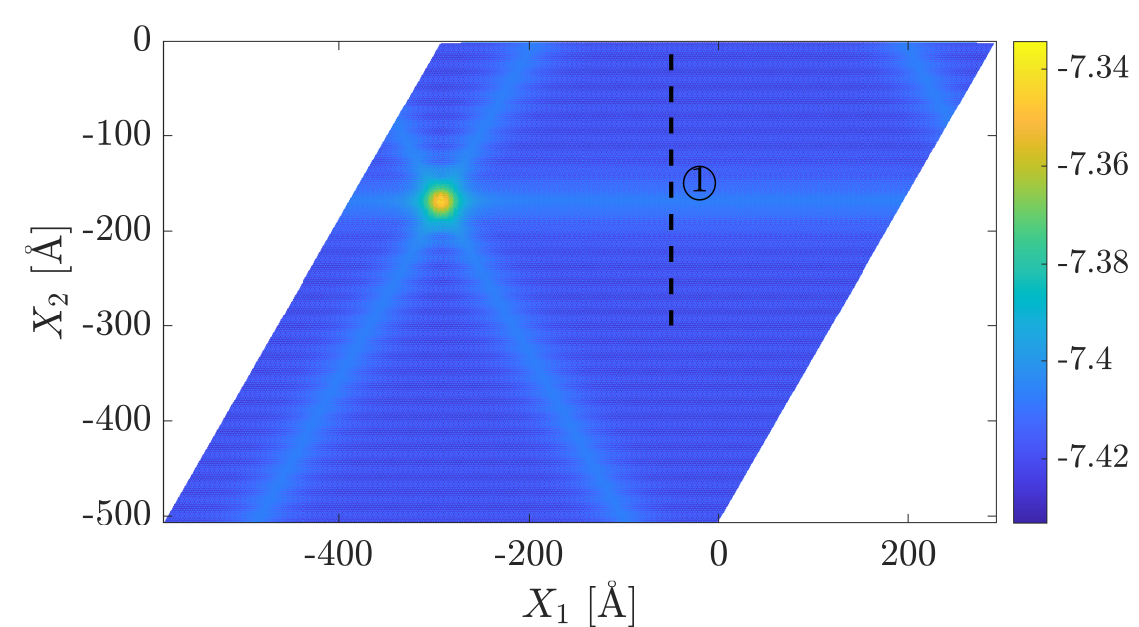}
        \label{fig:triangle_edge_contour}
    }
        \subfloat[configuration at step $659902$]
    {
        \includegraphics[height=0.25\textwidth]{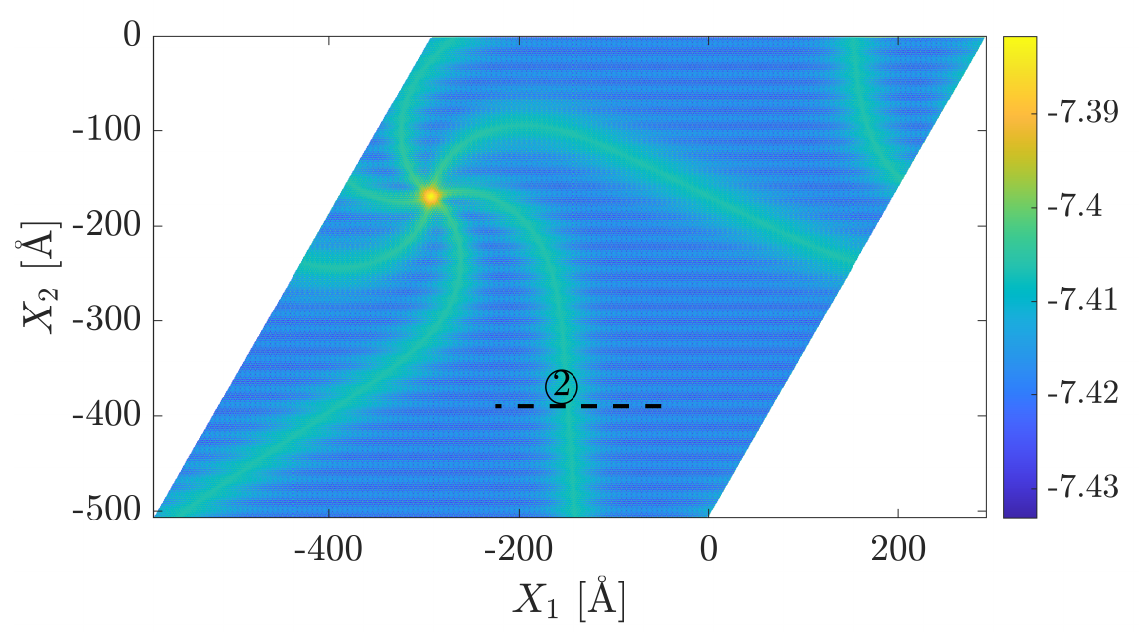}
        \label{fig:spiral_contour}
    }
    \\
    \subfloat[displacements at step $10000$]
    {
        \includegraphics[height=0.25\textwidth]{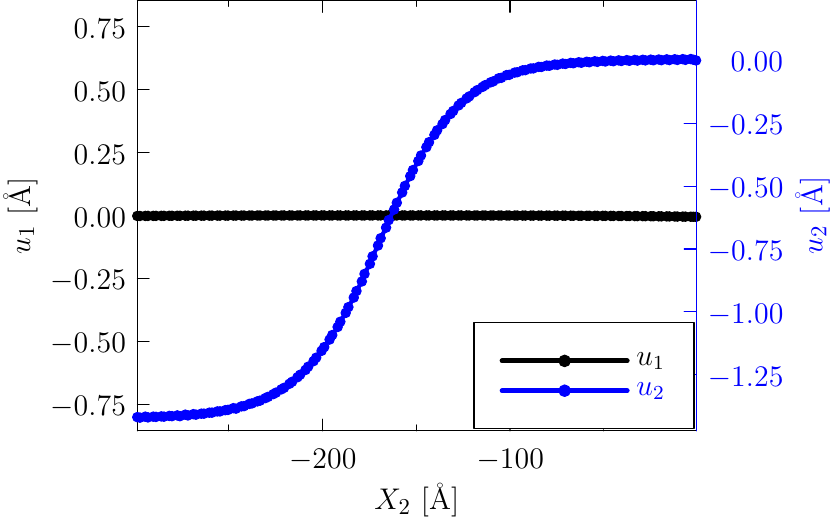}
        \label{fig:triangle_linescan}
    }
        \subfloat[displacements at step $659902$]
    {
        \includegraphics[height=0.25\textwidth]{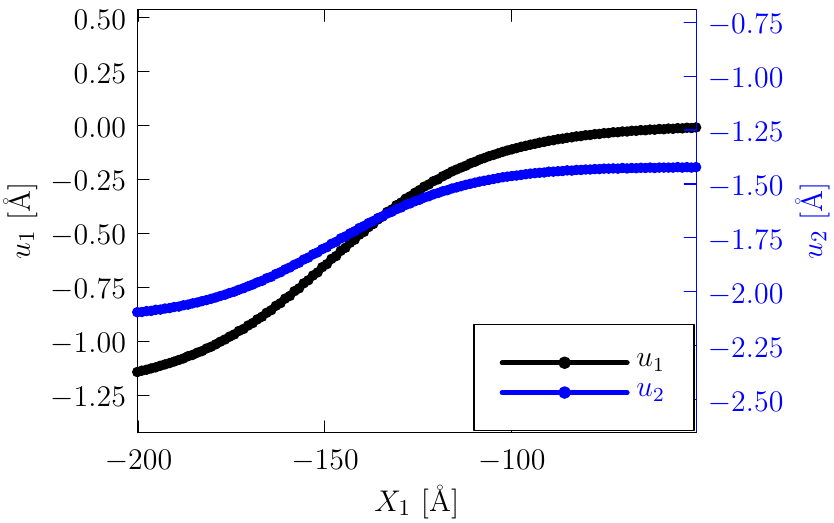}
        \label{fig:linescan_x_spiral}
    }
    \caption{Atomic reconstruction in $0.4219\%$ equi-biaxial heterostrained BG. \protect\subref{fig:energy_landscape} shows the evolution of the total energy during energy minimization. \protect\subref{fig:triangle_edge_contour} and \protect\subref{fig:spiral_contour} Atomic energy per atom [in $\si{\eV}$], observed at steps $10000$ and $~659902$ respectively, show the formation of a network of straight dislocations initially, followed by swirling dislocations at the end of minimization. \protect\subref{fig:triangle_linescan} and \protect\subref{fig:linescan_x_spiral} Displacement [in $\si{\angstrom}$] plots along lines 1 and 2 show that the swirling transforms the initial edge character of the dislocations to a mixed character.
    }
\label{fig:edge_atomistics}
    \end{figure}
Moving on to the equi-biaxial heterostrain case, we expect the dislocations to have a pure edge character \citep{harley_disloc, Annevelink_2020}. While the structure passes through a configuration with edge dislocations at $\approx 10000$-th step (see \fref{fig:triangle_edge_contour}), the energy plot in \fref{fig:energy_landscape} suggests it is metastable. The straight edge dislocation lines eventually swirl at the end of the minimization, as shown in \fref{fig:spiral_contour}. \frefs{fig:triangle_linescan}{fig:linescan_x_spiral} show line scans of displacements --- measured relative to the AB stacking --- along lines 1 and 2, marked in the corresponding energy per atom plots. The displacement plots confirm that the swirling transforms the initial edge character of dislocations, with a Burger vector magnitude $1.42~\si{\angstrom}$, into a mixed character. We make the following observations:
\begin{enumerate}
    \item Since the energy of a screw dislocation is lower than that of an edge dislocation (screw $\rightarrow$ $0.04~\si{eV\angstrom}^{-1}$, edge $\rightarrow$ $0.11~\si{eV\angstrom}^{-1}$ in BG \citep{Annevelink_2020}), the dislocation lines swirl to lower their edge character and increase their screw character. As a result, the maximum energy per atom drops from $-7.33~\text{eV/atom}$ to $-7.37~\text{eV/atom}$. \citet{mesple2023giant} noted that the edge character decreases as the heterostrain decreases.
    \item While the dislocation swirl in \fref{fig:spiral_contour} has a counterclockwise chirality, it can be inferred \cite{mesple2023giant} by symmetry that an energetically equivalent configuration exists with a clockwise chirality.
\end{enumerate}

The two examples discussed in this simulation demonstrate that heterodeformation significantly influences the network of dislocations. The next section will explore how dislocations respond to external loads and influence interlayer friction.

\subsection{Dynamics of heterodeformed BG}
\label{sec:dynamics_atom}
In this section, we study the interfacial friction in a heterodeformed BG when subjected to shear stress. At the macroscale, we measure interfacial friction and its dependence on the twist angle, sliding velocity, and the normal load. In addition, we correlate friction to the motion of interface dislocations at the atomic scale.

\subsubsection{A macroscopic view of friction}
\label{sec:macro_char_dynamics}
The goal of this section is to investigate dynamic friction using MD simulations of a BG subjected to shear under an out-of-plane compression. In finite-sized BG, friction arises not only from the interface but also from the contact edges \citep{qu2020origin}. In this study, however, we limit ourselves to studying interfacial friction in systems without contact edges using periodic boundary conditions.

\begin{figure}
    \centering
    \includegraphics[height=0.35\textwidth]{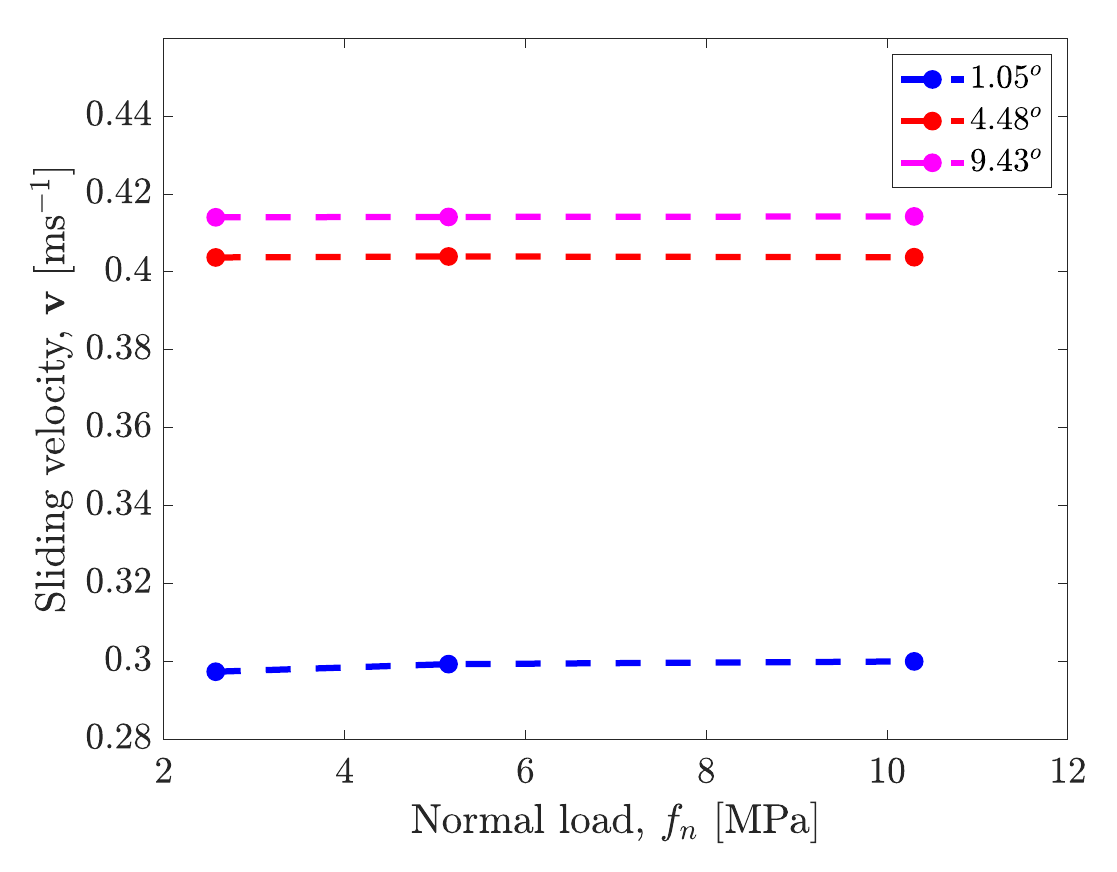}
    \caption{Variation of sliding velocity with the normal load in twisted BGs subjected to shear traction of $\SI{3.296}{\mega\pascal}$.}
    \label{fig:normal_twist}
\end{figure}

Classically, friction force $\bm f$ at the interface of two bulk solid materials originates at the contact points due to interface corrugation \citep{bowden2001friction} and is described by the phenomenological relation $f= \mu_{\rm k} f_{\rm n}$, where $f_{\rm n}$ is the magnitude of the normal force and $\mu_{\rm k}$ is the coefficient of kinetic friction. However, friction in BG is fundamentally different. Due to the atomically smooth nature of the BG interface, the dependence of friction on the normal load is insignificant \citep{lin2022friction,song2018robust,wang2024colloquium,song2022velocity}. To confirm the normal load independence of friction, we performed MD simulations of three BGs subjected to shear traction of $\SI{3.296}{\mega\pascal}$ at normal loads in the experimentally accessible range of $\SI{2}{\mega\pascal}$-$\SI{10}{\mega\pascal}$ \citep{qu2020origin}. \fref{fig:normal_twist} shows the variation of the sliding velocity $\bm v$ with the normal load $\bm f_{\rm n}$, and the three plots confirm that the normal load's influence on friction is insignificant. 

\begin{figure}
    \centering
    \subfloat[]
    {
    \includegraphics[height=0.35\textwidth]{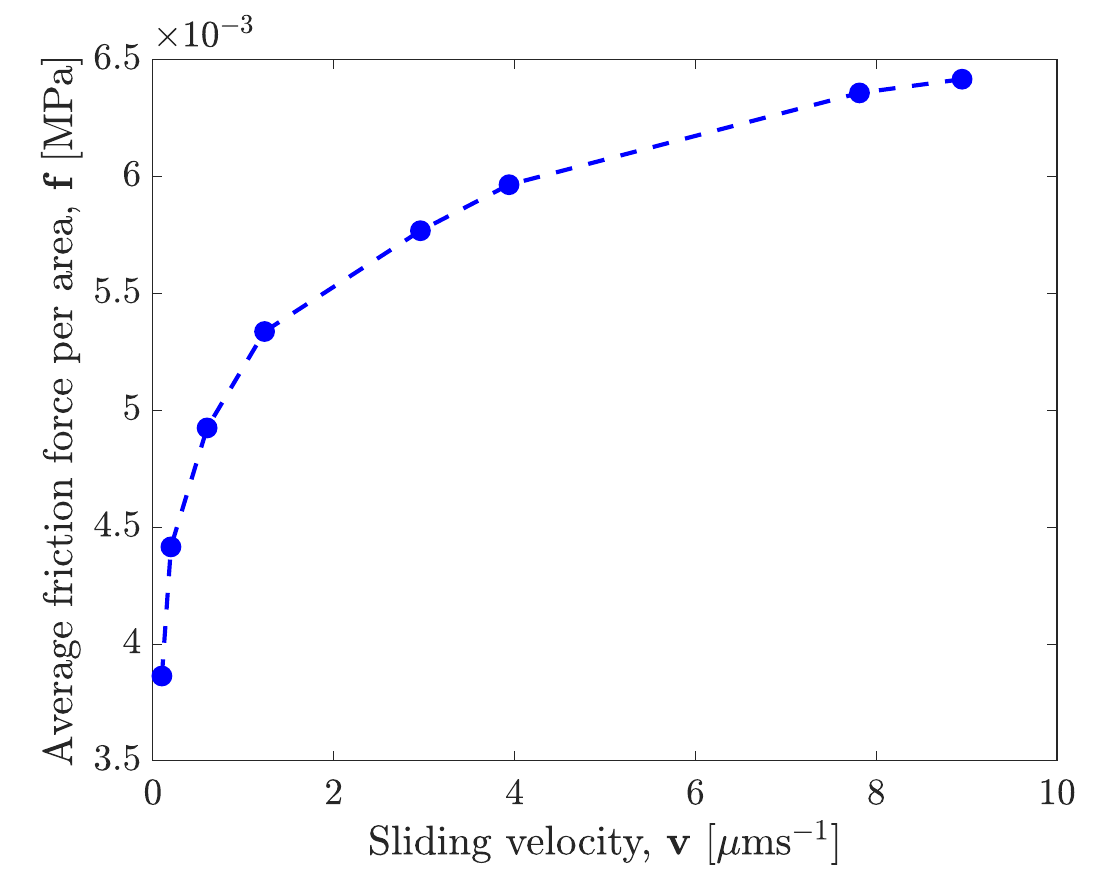}
        \label{fig:exp_plot}
    }
    \subfloat[]
    {
        \includegraphics[height=0.35\textwidth]{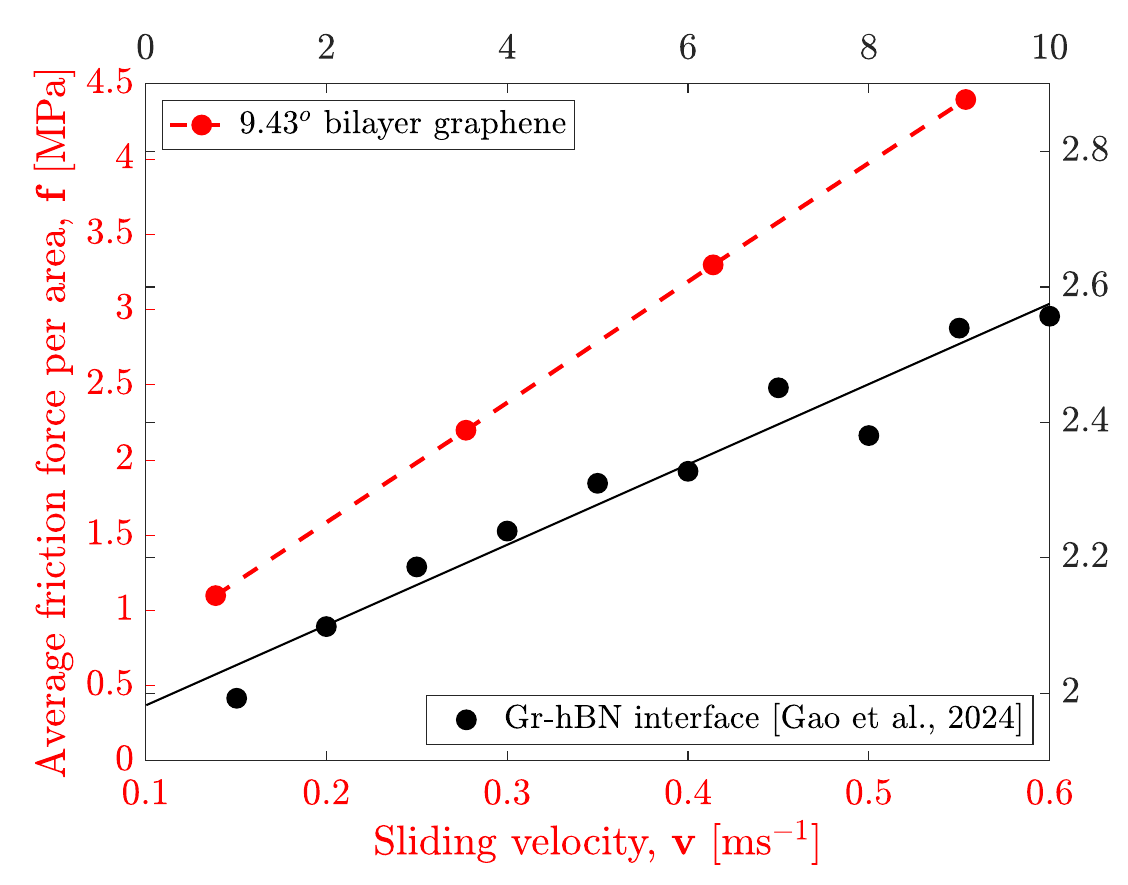}
        \label{fig:sim_plot}
    }
    \caption{Variation of friction force with sliding velocity. \protect\subref{fig:exp_plot} In the low sliding velocity regime, friction in a bilayer graphite interface (adapted from \citet{qu2020origin}), under a normal load of $\SI{2.375}{\mega\pascal}$, shows a logarithmic dependence on the sliding velocity. \protect\subref{fig:sim_plot} In a high-velocity regime, friction varies linearly with sliding velocity. 
    The data in black is adapted from \citet{gao2024dependency} for a graphene-hexagonal boron nitride interface simulated under an out-of-plane pressure of $10.280$ MPa, while the data in red is for a $9.43^{\circ}$ twisted BG under similar pressure, computed using LAMMPS.}
    \label{fig:literature_comp_plot}
\end{figure}

Next, we investigate the relationship between friction force and sliding velocity. Atomic force microscopy measurements by \citet{qu2020origin,song2018robust} show a logarithmic dependence of friction (see \fref{fig:exp_plot}) on the sliding velocity ranging from 0-$\SI{1}{\micro\meter\per\second}$. On the other hand, MD simulations \citep{ruan2021robust,gao2024dependency}, which cannot access the low experimental sliding velocities, predict a linear relationship between $\bm f$ and $\bm v$ when $v>\SI{1000}{\micro\meter\per\second}$. \fref{fig:sim_plot} showing our MD simulation data of a $\SI{9.43}{\degree}$ twisted BG (in red) and that of \citet{gao2024dependency} for a graphene-hexagonal boron nitride interface (in black) confirms the linear dependence of friction force in the high sliding velocity regime. Therefore, friction in 2D bilayer systems resembles the classical viscous drag observed in a fluid flowing over a solid surface \citep{wang2023kinetic} and satisfies the constitutive law 
\begin{equation}
    \bm f = \mu \bm v,
    \label{eqn:constitutive}
\end{equation}
where $\mu$ is the friction drag coefficient. Note that the constitutive law in \eqref{eqn:constitutive} ignores the directional anisotropy of the drag coefficient as we observe from our MD simulations that the sliding velocity is always parallel to the applied shear force in twisted and equi-biaxial heterostrained BGs. However, in a generic heterodeformed BG interface, the friction drag may have a tensorial character.
From \fref{fig:literature_comp_plot}, we also note that the drag coefficient in \eqref{eqn:constitutive} may depend on the sliding velocity and is independent of it only in the high-velocity regime.

\begin{figure}
    \centering
        \subfloat[]
    {
        \includegraphics[height=0.35\textwidth]{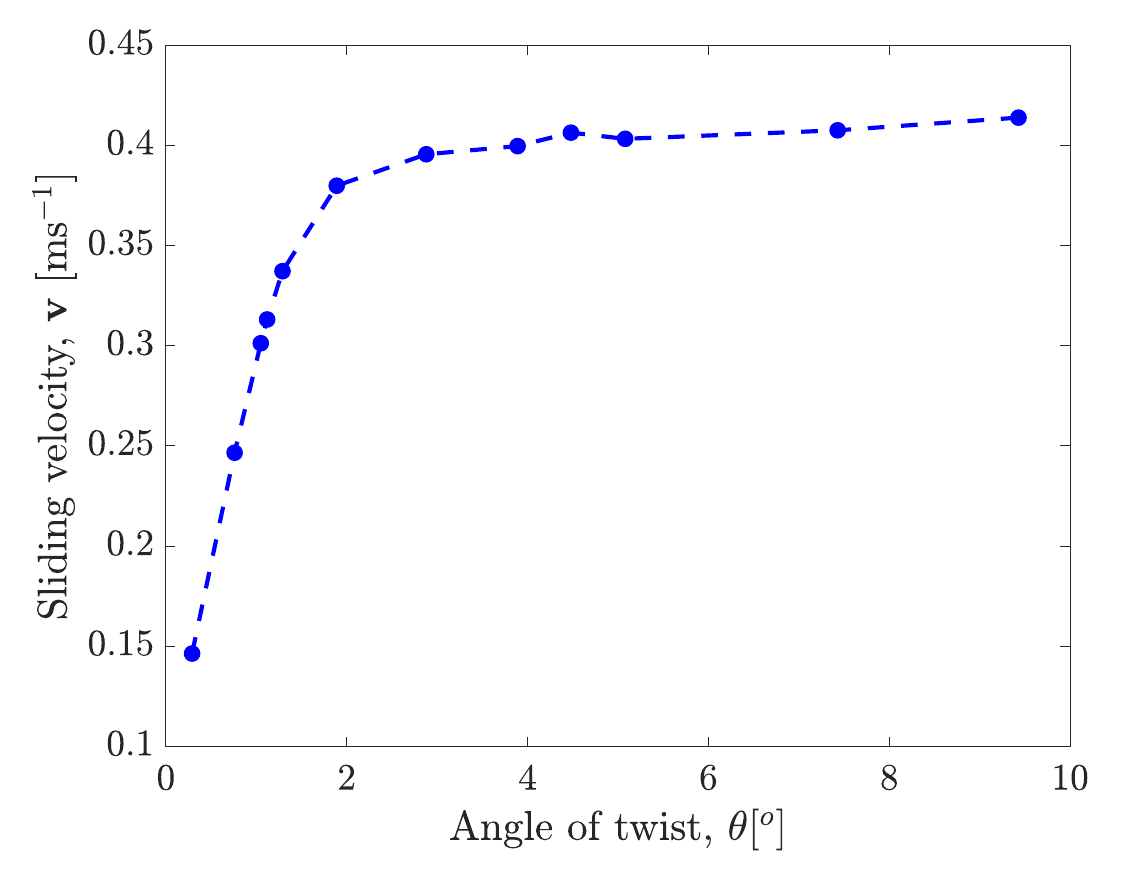}
        \label{fig:com_motion_atomistics}
    }
    \subfloat[]
    {
    \includegraphics[height=0.35\linewidth]{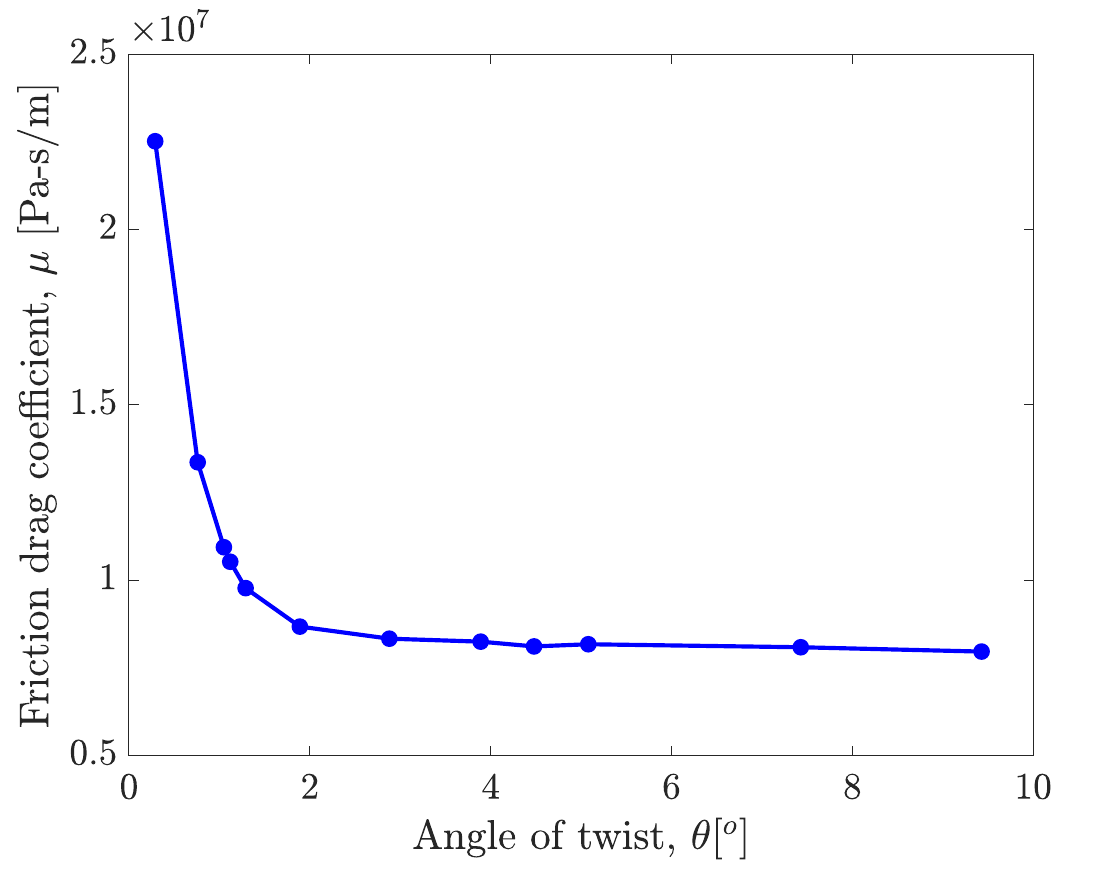}
    \label{fig:friction_atomistic}
    }
    \caption{Variation of \protect\subref{fig:com_motion_atomistics} sliding velocity and \protect\subref{fig:friction_atomistic} friction drag coefficient with the angle of twist in twisted BGs subjected to an out-of-plane compression of $\SI{2.57}{\mega\pascal}$ and shear traction of $\SI{3.296}{\mega\pascal}$.}
    \label{fig:macro_par_BG}
\end{figure}

Finally, we inspect the dependence of $\mu$ on the twist angle. \fref{fig:com_motion_atomistics} shows the variation of sliding velocity with the twist angle. Since the sliding velocity increases with the twist angle, it follows from \eqref{eqn:constitutive} that $\mu$, plotted in \fref{fig:friction_atomistic}, decreases as the twist angle increases. Moreover, the decrement in $\mu$ is rapid at small twist angles ($\theta < 2^{\circ}$). 
The macroscopic characteristics of friction noted in this section and the interface dislocations described in \sref{sec:atomic_recons} motivate us to inspect the dynamics of interface dislocations at the atomic scale and their influence on friction.

\subsubsection{Friction at the microscale}
\label{sec:microscale_char}
\begin{figure}
    \centering
    \subfloat[$t=0$]
    {
        \includegraphics[height=0.25\textwidth]{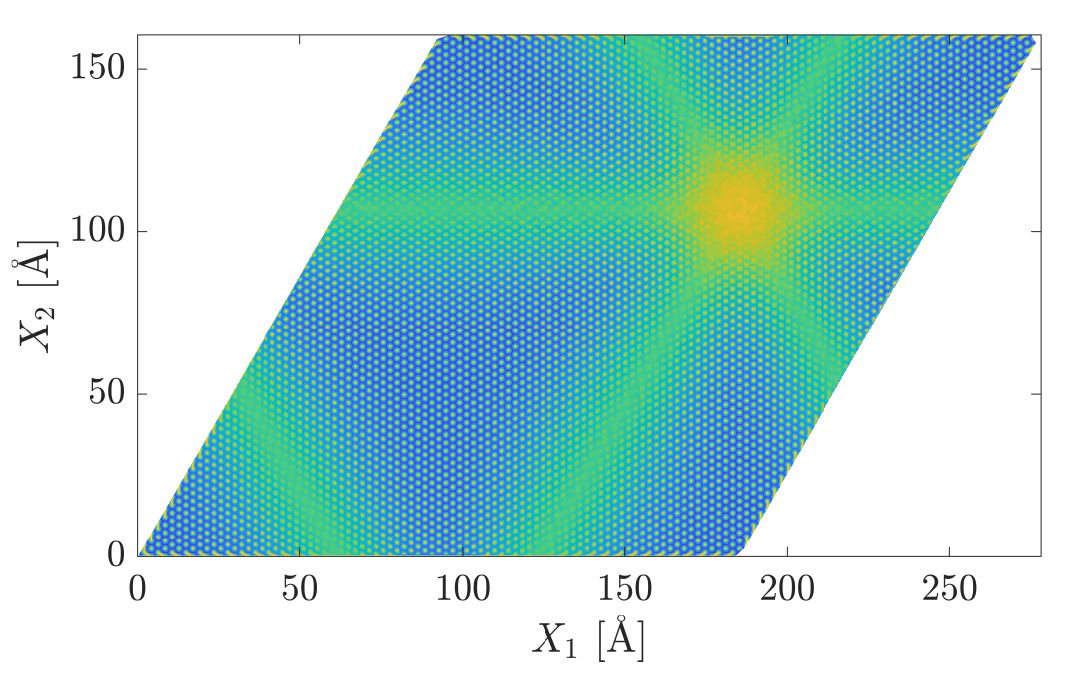}
        \label{fig:slide_initial}
    }
    \subfloat[$t=\SI{0.3}{\nano\second}$]
    {
        \includegraphics[height=0.25\textwidth]{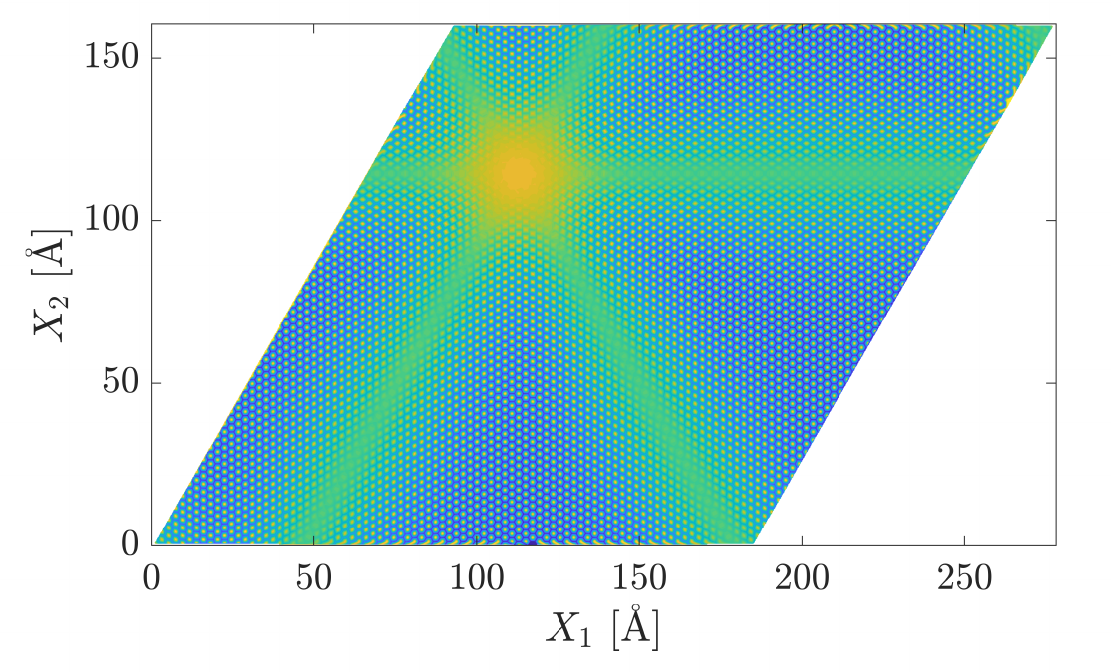}
        \label{fig:slide_final}
    }
    \caption{The translation of the dislocation network in a $0.76^{\circ}$ twisted BG when subjected to a shear traction along the $X_2$ direction. \protect\subref{fig:slide_initial} and \protect\subref{fig:slide_final} show the configuration before and after the application of shear traction, respectively. The simulation movie is provided as supplementary material.}
    \label{fig:slide_result}
\end{figure}

In this section, we study friction at the microscale by inspecting the response of interface dislocations in MD simulations of BG subjected to shear. \fref{fig:slide_initial} shows a structurally relaxed $0.760443^\circ$ twisted BG with a triangular network of screw dislocations, taken as the initial configuration of the MD simulation. When subjected to a shear stress of $\sigma_{23}= \SI{3.296}{\mega\pascal}$, the two layers start sliding with a constant velocity. During sliding, the cores of the dislocations remain compact while the network translates in unison in the $X_1$ direction, i.e., \emph{orthogonal} to the direction of the shear traction. \fref{fig:slide_final} shows the configuration at $t=\SI{0.3}{\nano\second}$.

\begin{figure}
    \centering
    \subfloat[$t=0$]
    {
        \includegraphics[height=0.25\textwidth]{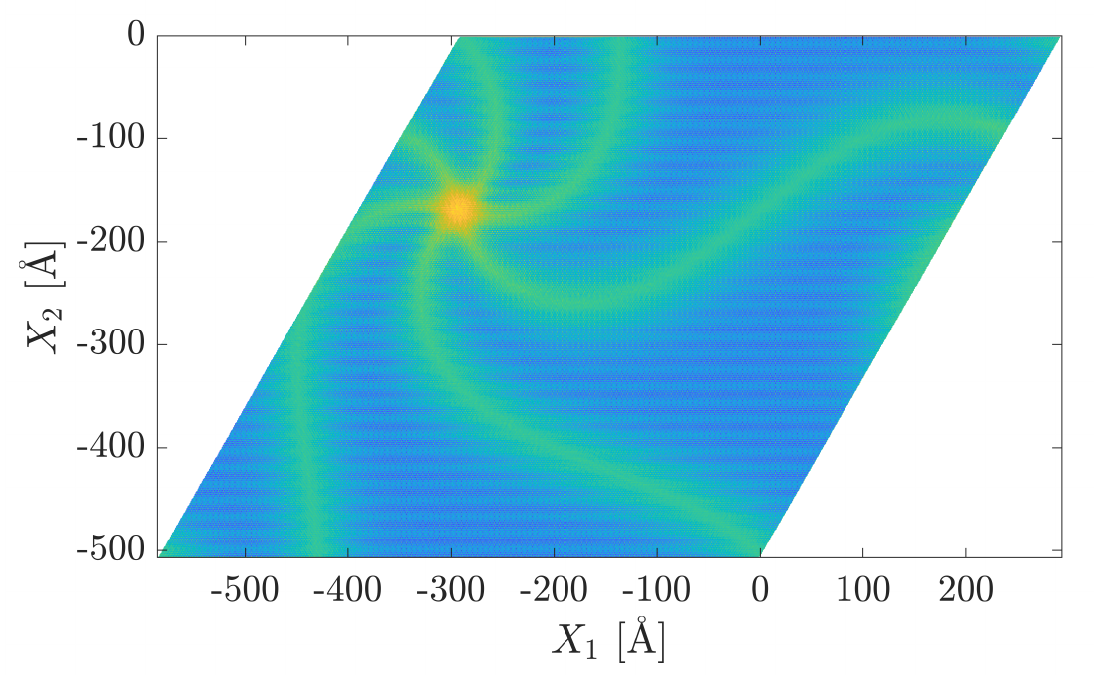}
        \label{fig:slide_initial_str}
    }
    \subfloat[$t=\SI{1}{\nano\second}$]
    {
        \includegraphics[height=0.25\textwidth]{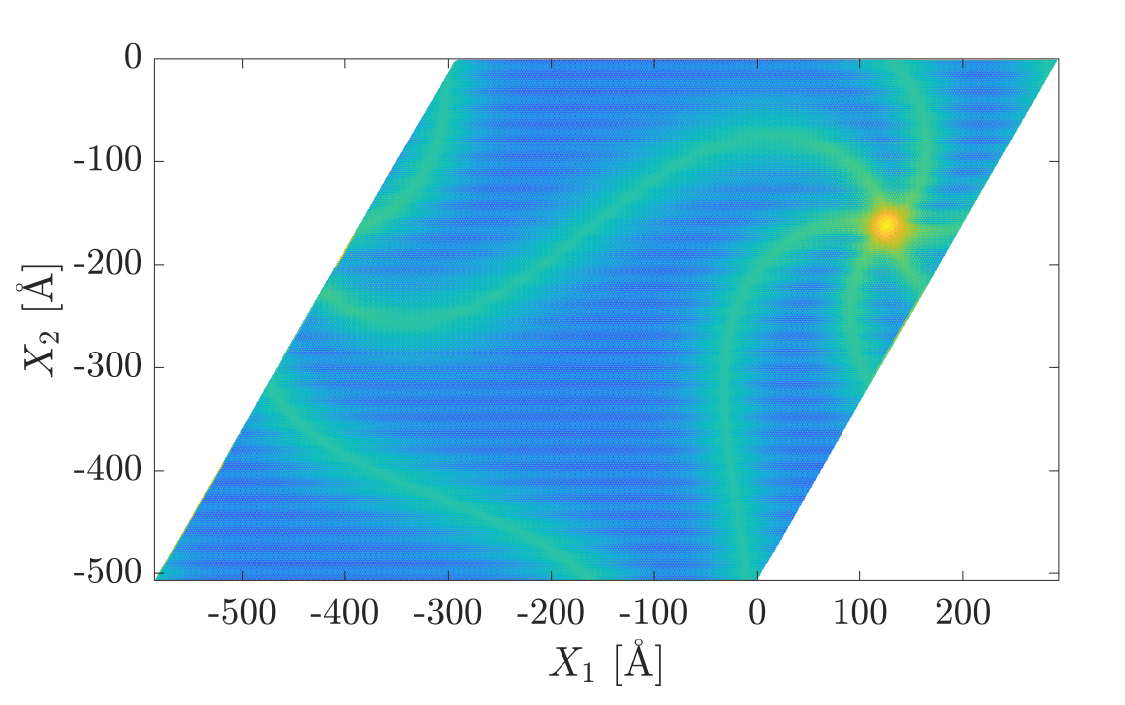}
        \label{fig:slide_final_str}
    }
    
    \caption{The translation of the dislocation network in a $0.4219\%$ equi-biaxial heterostrained BG when subjected to shear traction along the $X_1$ direction. \protect\subref{fig:slide_initial_str} and \protect\subref{fig:slide_final_str} show the configuration before and after the application of shear traction, respectively. The simulation movie is provided as supplementary material}
    \label{fig:slide_result_str}
\end{figure}

Next, we inspect the response of interface dislocations with spatially varying character in the $0.4219\%$ equi-biaxial heterostrained BG introduced in \sref{sec:atomic_recons}. When subjected to a shear stress of $\sigma_{13}=\SI{3.296}{\mega\pascal}$, the swirling dislocations translate in unison in the $X_1$ direction, i.e., \emph{parallel} to the direction of shear, as shown in \fref{fig:slide_result_str}. While the interface dislocations possess an average edge character determined by the heterodeformation, it is remarkable that the spatially non-uniform dislocation character does not impede the uniform translation of the network. The above two examples highlight how the average character (screw/edge) of dislocations determines the direction (orthogonal/parallel) of the network translation as also noted by \citet{huang2024general}.

\begin{figure}
    \centering
    \includegraphics[height=0.35\textwidth]{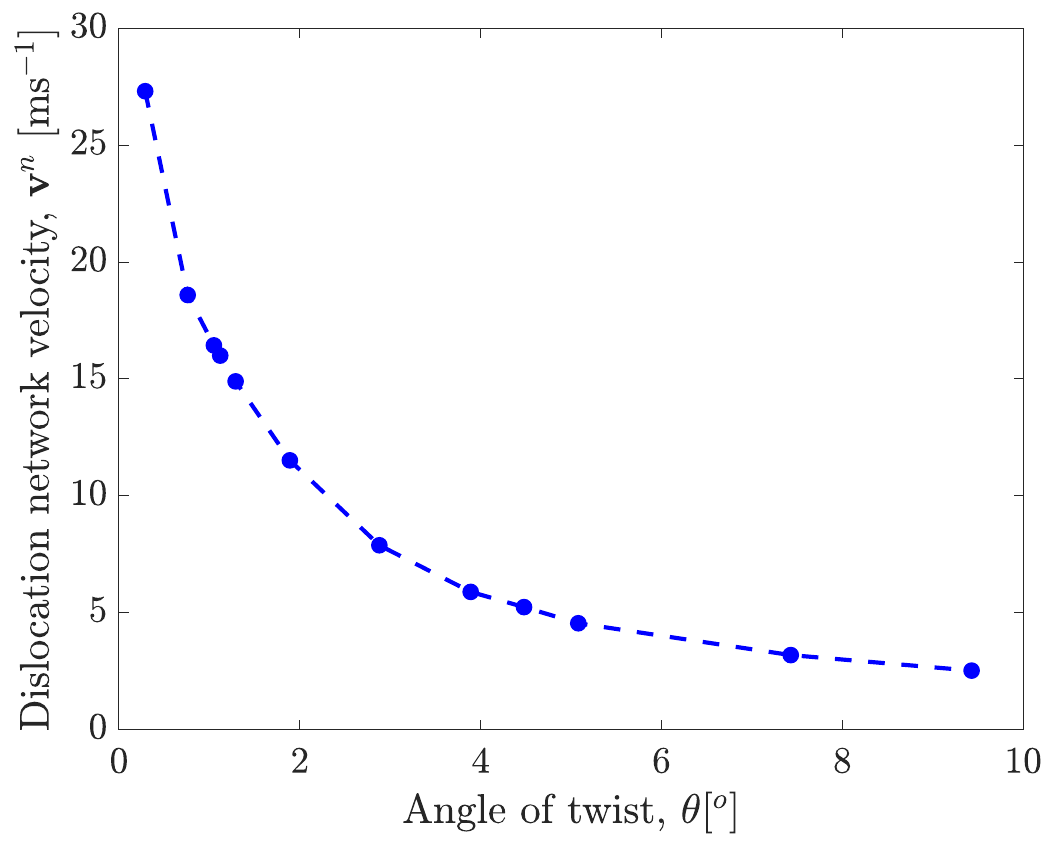}
    \caption{Variation of the network velocity with the angle of twist in twisted BGs.}
    \label{fig:moire_motion_atomistics}
\end{figure}

From a defects viewpoint, the interface dislocations move in response to the Peach--Koehler force originating from the applied shear, and the steady-state velocity $\bm v^{\rm n}$ of the network is determined by their mobility. Therefore, defect mechanics relates the applied force to the network velocity. However, to arrive at the friction drag coefficient, we need to connect the network velocity to the sliding velocity. It is interesting to note that compared to the increasing sliding velocity with the twist angle, observed in \fref{fig:com_motion_atomistics}, the network velocity magnitude $v^{\rm n}$ (see \fref{fig:moire_motion_atomistics}) follows the opposite trend, i.e. $v^{\rm n}$ decreases as the twist angle increases. 
The relationship between the sliding velocity and the network velocity is known to be geometric in nature. Using the framework of SNF bicrystallography, \citet{ADMAL2022} derived the relationship $\bm v^{\rm n}=\bm \Lambda \bm v$, where $\bm \Lambda$ is a heterodeformation-dependent tensor.\footnote{See \citet{Hermann_2012} and \citet{ADMAL2022} for derivations specialized to twists and equi-biaxial heterostrained cases.}  The above observations lead us to the following assertion:
\begin{quote}
    The friction drag coefficient as a function of the four-dimensional heterodeformation space can be determined entirely by the character-dependent mobility of dislocations.
\end{quote}
To demonstrate the validity of the above assertion, in the next section, we develop a continuum model based on the generalized Frenkel--Kontorova model of \citet{ahmed2024bicrystallography}.

\section{Continuum model to predict interface friction}
\label{sec:cont_fram}
In this section, we formulate a dynamic Frenkel--Kontorova (DFK) model to predict friction in heterodeformed BG. The model is a dynamic version of the structural reconstruction mesoscale model of \citet{ahmed2024bicrystallography}. In particular, the goal is to predict friction drag coefficient for any heterodeformation. A key feature of the model is frame-invariance, which enables it to model large heterodeformations. For simplicity, the model ignores the out-of-plane displacement caused by interface dislocations.

\subsection{Kinematics}
The DFK model is designed to predict the two time-varying displacement fields of the two layers of a heterodeformed BG subjected to shear stress. The displacement fields are measured relative to a uniformly heterodeformed (relative to the AB stacking) BG, which is chosen as a reference configuration. Therefore, unlike in classical continuum models, the reference configuration is not strain-free. To account for the elastic energy and the vdW interaction energy, strain-free AB stacking is introduced as an intermediate/natural configuration. 

The two layers of a BG are modeled as 2D continua in the 2D Euclidean point space $\mathbb R^2$, and the corresponding uniformly heterodeformed reference configurations are denoted by $\Omega^{\rm ref}_{\rm t}$ and $\Omega^{\rm ref}_{\rm b} \subset \mathbb R^2$.\footnote{The subscripts $\rm t$ and $\rm b$ indicate top and bottom layers, respectively.} An arbitrary material point in the reference configuration is denoted by $\bm X_\alpha$ ($\alpha=\rm t,\rm b$). Time-dependent displacement maps $\bm \phi_\alpha:\Omega^{\rm ref}_\alpha \times [0,\infty] \to \mathbb R^2$ of the two layers describe the dynamics of structural relaxation and sliding associated with friction. Therefore, 
$\bm \phi_\alpha-\bm X_\alpha$ describes the displacement of the $\alpha$-th layer relative to the reference configuration.

To construct the elastic and vdW energies in \sref{sec:constitutive}, we introduce the AB-stacked intermediate configuration and a map $\bm \kappa_\alpha$, which maps the reference to the natural configuration. For example, in the context of the two examples introduced in \sref{sec:friction}, $\bm \kappa_{\rm t}(\bm X_{\rm t})= \bm X_{\rm t}$ and $\bm \kappa_{\rm b}(\bm X_{\rm b})= \bm F^{-1}\bm X_{\rm b}$. Therefore, $\bm \kappa_\alpha$ is prescribed by the imposed heterodeformation. A mapping $\bm \eta_\alpha$ from the intermediate to the deformed configuration is defined such that  
\begin{equation}
    \bm \phi_\alpha = \bm \eta_\alpha \circ \bm \kappa_\alpha,
    \label{eqn:composition}
\end{equation}
where $\circ$ denotes function composition. The deformation gradients corresponding to the deformation maps are given by
\begin{equation}
    \bm F_\alpha:=\nabla \bm \phi_\alpha = \bm H_\alpha \bm K_\alpha, \text{ where } \bm H_\alpha:=\nabla \bm \eta_\alpha \text{ and, } \bm K_\alpha:=\nabla \bm \kappa_\alpha .
   \label{eqn:FeFp}
\end{equation}
Since the map $\bm \eta_\alpha$ describes elastic distortions relative to the intermediate configuration, the frame invariant kinematic measure $\bm C_\alpha:= \bm H_\alpha^{\rm T} \bm H_\alpha$ will be used to construct the elastic energy in \sref{sec:constitutive}. From \eqref{eqn:FeFp}, $\bm C_\alpha$ can be written as
\begin{equation}
    \bm C_\alpha = \bm K_\alpha^{-\rm T} \bm F_\alpha^{\rm T}\bm F_\alpha \bm K_\alpha^{-1}.
\end{equation}
On the other hand, the vdW interfacial energy depends on the lattice stacking, which is characterized by the relative translation between the two layers. Thus, the interfacial energy density at any point $\bm x$ in the deformed configuration is determined by the relative translation vector
\begin{equation} 
\bm r(\bm x,t)= \Kt \Xt-\Kb \Xb, \text{ where }
    \bm X_\alpha:= \bm \phi_\alpha^{-1}(\bm x,t).
    \label{eqn:r}
\end{equation}
The next section will use the frame-invariant kinematic measures, $\bm C_\alpha$ and $\bm r$, to construct constitutive laws for the elastic and interfacial energies, respectively.
\subsection{Constitutive law}
\label{sec:constitutive}
The total energy of a heterodeformed BG subjected to shear is composed of elastic energy, interfacial energy, and the work done due to external force, i.e.,  $\mathcal E=\mathcal E_{\rm el}+\mathcal E_{\rm vdW}+\mathcal E_{\rm external}$ \citep{Koshino_2017}.
The elastic energy is given in terms of the frame-invariant $\bm C_\alpha$ as
\begin{equation}
    \mathcal E_{\rm el}[\phit, \phib] = \sum_{\alpha=\mathrm t, \mathrm b}
    \int_{\Omega_\alpha^{\rm n}} 
    e_{\rm el}(\bm E_\alpha;\alpha) \, d\bm Y_\alpha, \quad  \quad \text{where } \bm E_{\alpha}=(\bm C_\alpha-\bm I)/2
    \label{eqn:elasticEnergy}
\end{equation}
is the frame-invariant Lagrangian elastic strain tensor. Note that the integral in \eqref{eqn:elasticEnergy} is over the natural configuration, $\Omega^{\rm n}_\alpha$, whose points are denoted by $\bm Y_{\alpha}$. Moreover, $e_{\rm el}(\bullet; \alpha)$ refers to the elastic energy density of the $\alpha$-th layer, and is assumed to be of the Saint Venant–Kirchhoff type:
\begin{equation}
    e_{\rm el}(\bm E_\alpha;\alpha) = \frac{1}{2} \mathbb C \bm E_\alpha \cdot \bm E_\alpha= \lambda (\mathrm{tr}\, \bm E_\alpha)^2 + 2\mu \bm E_\alpha \cdot \bm E_\alpha,
\end{equation}
where $\mathbb C$ is the fourth-order isotropic elasticity tensor with lam\'e constants $\lambda$ and $\mu$. Isotropic lam\'e constants of graphene, as obtained from \citet{zakharchenko2009finite}, are $\lambda=\SI{3.25}{\eV\per\angstrom\squared}$ and $\mu=\SI{9.57}{\eV\per\angstrom\squared}$.

The vdW interfacial energy is constructed as the following integral over a region in the deformed configuration where the two layers overlap:
\begin{equation}
\mathcal E_{\rm vdW}[\phit, \phib] =  
\frac{1}{2}
\sum_{\alpha=\rm t, \rm b}
\int_{\Omega_{\rm t} \cap \Omega_{\rm b}} 
(\det \bm H_\alpha)^{-1} e_{\rm vdW}(\bm r(\bm x_\alpha)) \, d\bm x_\alpha,
\label{eqn:interEnergy}
\end{equation} 
where $\bm x_\alpha$ denotes an arbitrary point in the deformed configuration, and  $e_{\rm vdW}$ is the interfacial energy density, also referred to as the GSFE.
Note that the factor $(\det \bm H_\alpha)^{-1}$ is necessary because the integration is over the deformed configuration as opposed to the intermediate configuration.

The GSFE is a periodic function given by
\begin{equation}
    e_{\rm vdW}(\bm r) = \pm 2 v_0\sum_{p=1}^{3}\cos \left (2 \pi 
        \bm{\mathcal d}^p \cdot (\bm r+\bm s)
    \right)+c,
    \label{eq:vdW_en}
\end{equation}
where $v_0$ is the strength of the GSFE, $\bm{\mathcal d}^1$ and $\bm{\mathcal d}^2$ are primitive reciprocal vectors of the lattice,\footnote{Note that the two graphene lattices in the AB-stacking share a common lattice while their basis atoms differ.} and $\bm{\mathcal d}^3=-(\bm{\mathcal d}^1+\bm{\mathcal d}^2)$.  The constant $v_0$ is obtained by comparing the GSFE plot calculated using LAMMPS (shown in \fref{fig:GSFE_atom}) with \eqref{eq:vdW_en}. The constant vector $\bm s$ is used to position the GSFE such that its origin corresponds to the AB stacking while the scalar $c$ is a constant to change the limits of the GSFE for convenient comparison with atomistics. By comparing equation \ref{eq:vdW_en} with \fref{fig:GSFE_atom}, we have $v_0=0.00025~\si{eV\angstrom}^{-2}$, $\bm s=(-1.230 \bm e_1 - 0.710 \bm e_2)$\AA, and $c=7.5\times 10^{-4}~\si{eV\angstrom}^{-2}$.

Lastly, the contribution to the total energy due to the work done by applied traction $\bm t_\alpha$ is given by 
\begin{equation}
\mathcal E_{\rm external}[\phit, \phib] =  
- \sum_\alpha \int_{\Omega_\alpha}\bm t_\alpha \cdot \bm \phi_\alpha \, d\bm X_\alpha.
\label{eqn:extern_work}
\end{equation} 
In this work, we assume $\bm t_\alpha$s are constant vectors, and since the total applied force is zero, $\bm t_{\rm t}=-\bm t_{\rm b}=:\bm t$.

\subsection{Governing equations}
We assume the system evolves according to the gradient flow:
\begin{equation}
    b \dot{\bm \phi}_\alpha = -\updelta_{\bm \phi_\alpha} \mathcal E, \quad \alpha = \rm t, \rm b
    \label{eqn:gradientFlow}
\end{equation}
where, $b$ is the inverse mobility and $\updelta_{\bm \phi_\alpha}$ denotes variation with respect to $\bm \phi_\alpha$. Calculating the variational derivatives of \eqref{eqn:elasticEnergy}, \eqref{eqn:interEnergy}, and \eqref{eqn:extern_work}, and substituting them in \eqref{eqn:gradientFlow}, we obtain
\begin{subequations}
\begin{align}
    b \dot{\bm \phi}_{\rm t} &= \divrt(\bm P_{\rm t}) + \Hb^{-\rm T} \nabla e_{\rm vdW}(\bm r_{\rm t})+\bm t, \\
    b \dot{\bm \phi}_{\rm b} &= \divrb(\bm P_{\rm b}) -  \Ht^{-\rm T} \nabla e_{\rm vdW}(\bm r_{\rm b}) - \bm t,
\end{align}
\label{eq:final_eqn}%
\end{subequations} 
where $\bm P_\alpha:= \bm H_\alpha \bm \nabla \bm e_{\rm el} \bm K_\alpha^{-\rm T}$ is the 2D analog of the elastic Piola--Kirchhoff stress, which describes force (in $\Omega_\alpha$) per unit length in $\Omega_\alpha^{\rm ref}$.
The relative translation vectors, $\rt$ and $\rb$, can be expressed in terms of the unknown fields $\phit$ and $\phib$ using the approximation\footnote{See Section 4.3 in \citep{ahmed2024bicrystallography} for the fully general governing equations and the argument for the approximations in \eqref{eqn:r_approx}, in addition to boundary conditions relevant for finite systems.} 
\begin{subequations}
\begin{align}
    \rt &\approx (\Kt -\Kb) \Xt -  \Hb^{-1} (\phit(\Xt)-\phib(\Xt)),\label{eqn:rt_approx}\\
    \rb &\approx (\Kt -\Kb) \Xb -  \Ht^{-1} (\phit(\Xb)-\phib(\Xb))). \label{eqn:rb_approx}
\end{align}
\label{eqn:r_approx}%
\end{subequations} 
Here, we do not list the boundary conditions as we impose PBCs to avoid edge effects in the current study. Further details on the numerical implementation of the DFK model with PBCs are given in Section S$-2$ of supporting information.

At this point, the inverse mobility $b$ is the only model parameter that has to be determined. In the next section, we describe a procedure to determine $b$ from atomistic simulation. Subsequently, we will use the DFK model to predict friction in a heterodeformed BG. 

\section{Quantifying interface friction from dislocation mobility}
\label{sec:det_mobility}
The goal of this section is to demonstrate that a single kinetic parameter $b$ of the DFK model characterizes friction in any heterodeformed BG. We first obtain $b$ by simulating a moving dislocation using the DFK model and atomistics and comparing their predictions. Subsequently, we use the fitted DFK model to predict friction and microstructural changes during sliding. All MD simulations presented in this section are preceded by structural relaxation using the FIRE algorithm and are NVT ensembles at \SI{0.01}{\kelvin}.

\subsection{An atomistic dislocation dipole simulation to calculate inverse mobility $b$}
The goal of this section is to obtain the inverse mobility of the DFK model ($b$ in \eqref{eqn:gradientFlow}) using an atomistic simulation of a screw dislocation motion in a BG. As opposed to a single dislocation, we introduce a dipole of two screw partials due to PBCs that are enforced to avoid edge boundary effects.

\begin{figure}
    \centering
    \subfloat[]
    {
        \includegraphics[height=0.3\textwidth] {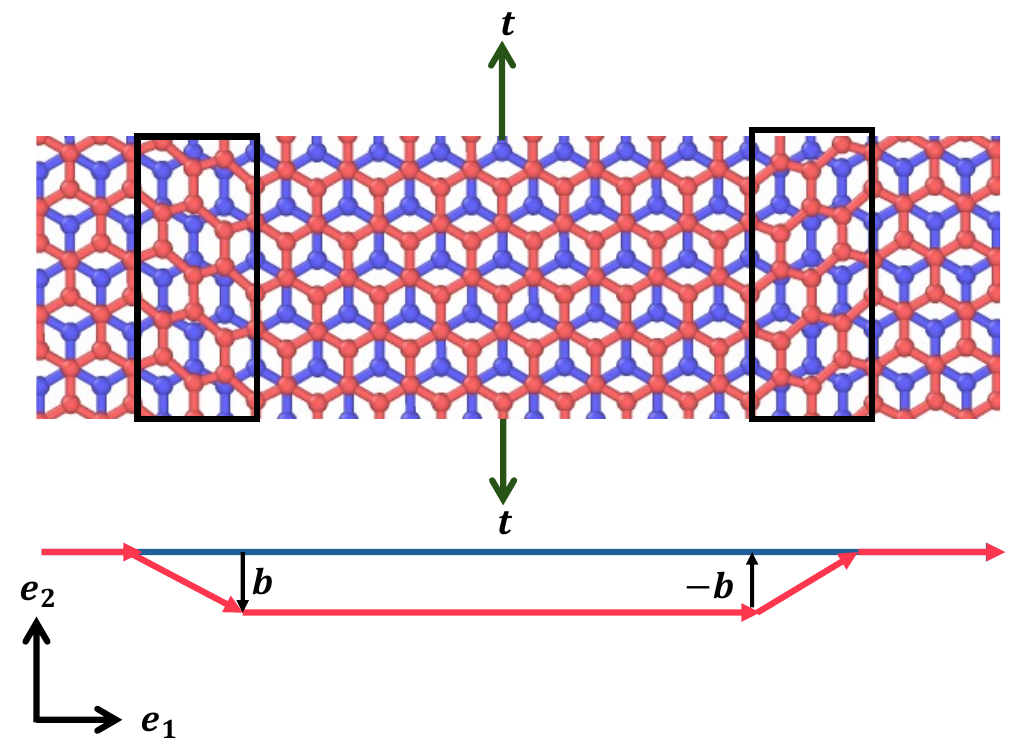}  \label{fig:dipole_schematic_atom}
    }
    \subfloat[]
    {
        \includegraphics[height=0.4\textwidth]{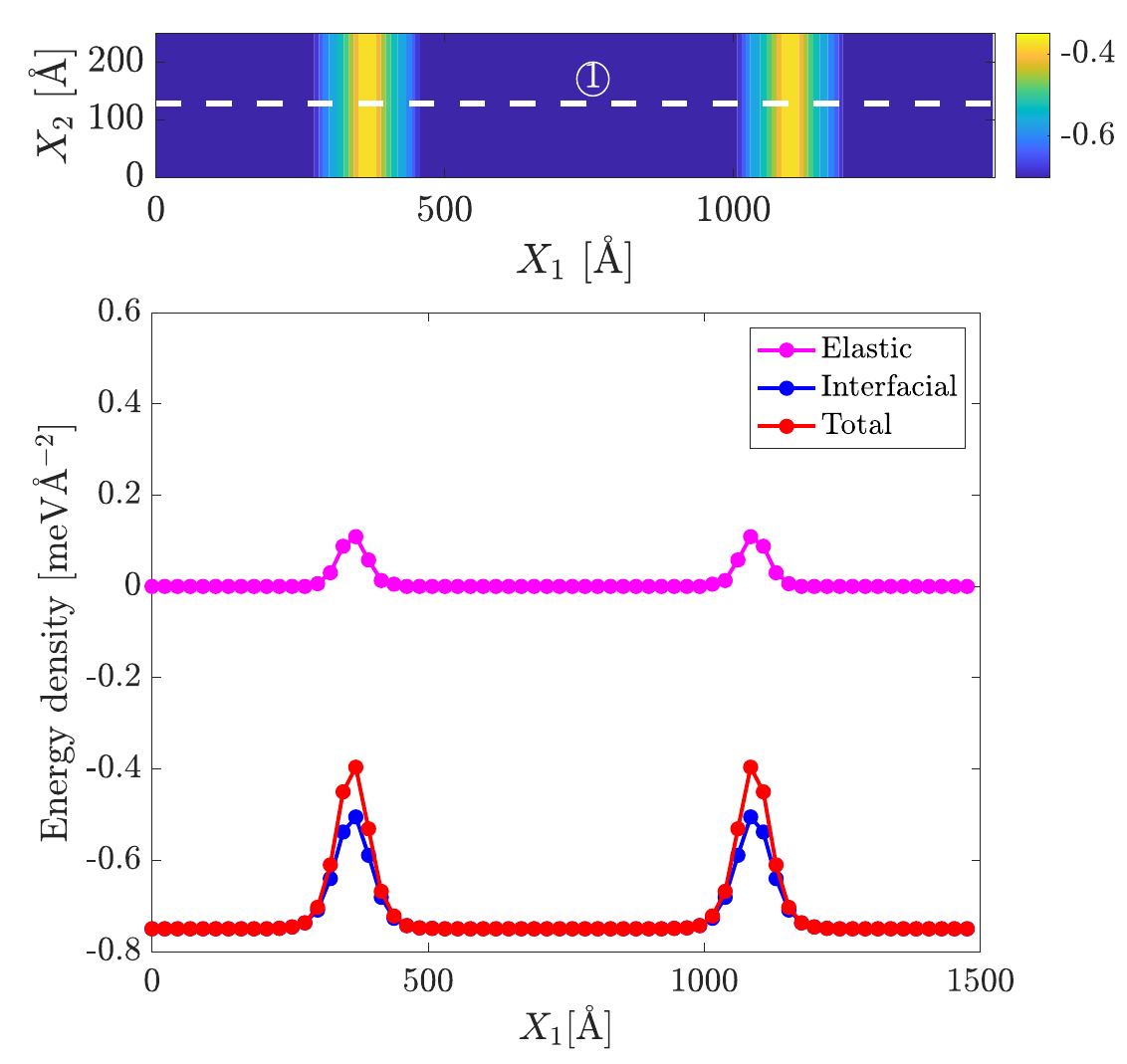}  \label{fig:dipole_schematic_cont}
    }\\
    \subfloat[]
    {
        \includegraphics[height=0.35\textwidth]{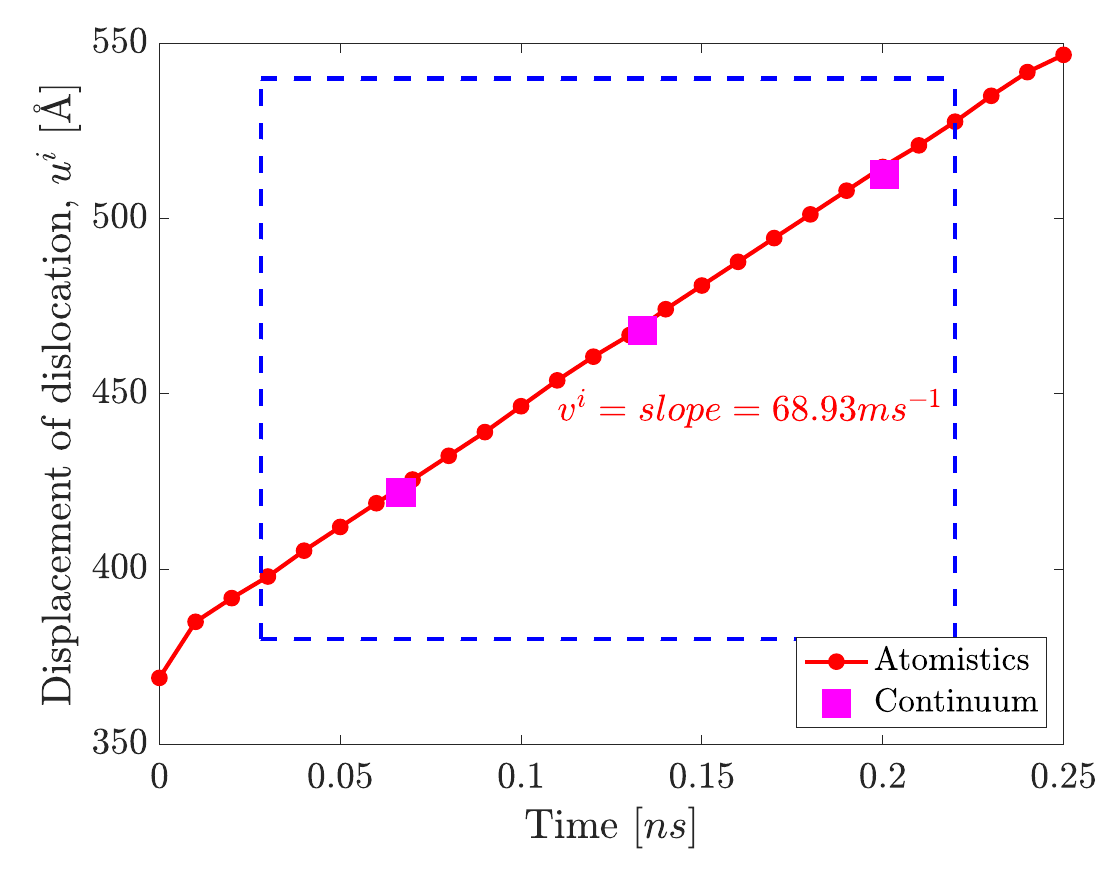}
        \label{fig:dipole_motion_comp_screw}
    }
    \subfloat[]
    {
        \includegraphics[height=0.35\textwidth]{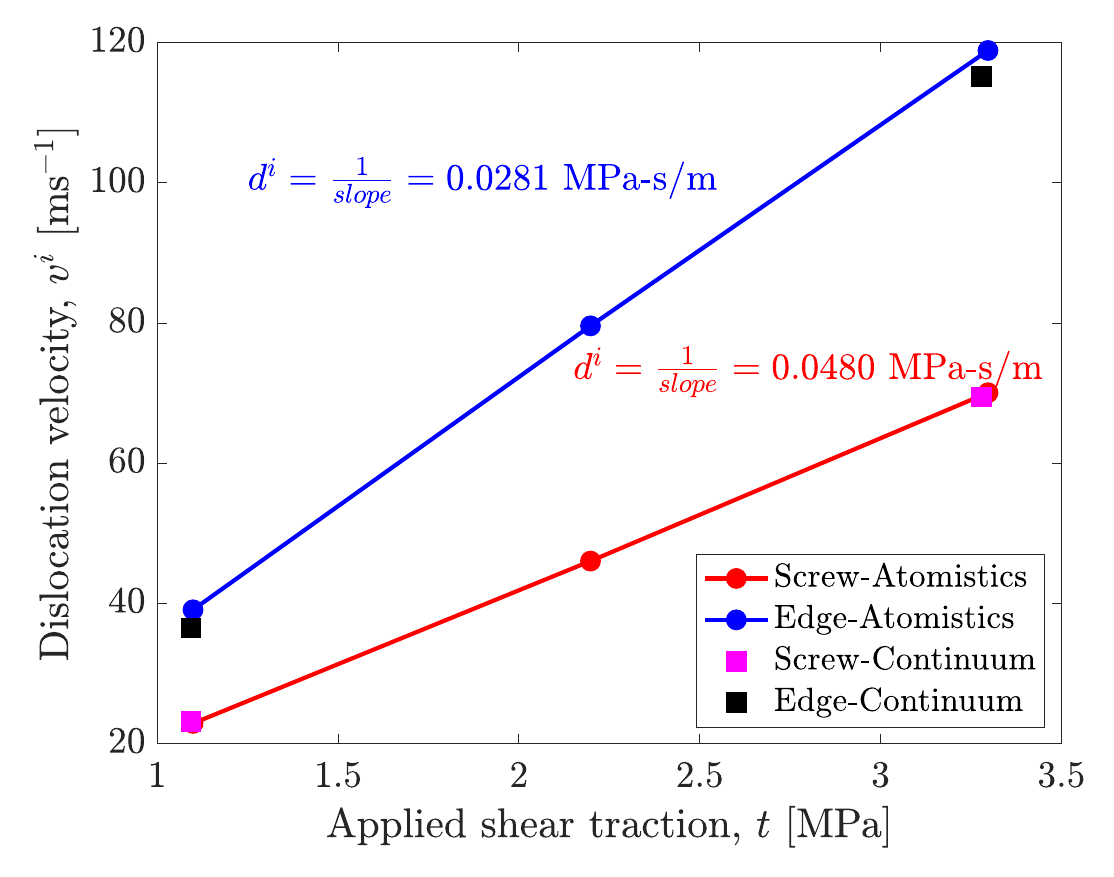}
        \label{fig:dipole_mobility_atom}
        }
    \caption{Atomistic and DFK model simulations of expanding dislocation dipole. \protect\subref{fig:dipole_schematic_atom} shows a schematic of a screw dislocation dipole in a BG along with the displacement used to generate the dipole. \protect\subref{fig:dipole_schematic_cont} shows the energy density [in $\si{\meV\angstrom}^{-2}$] plot of the equivalent screw dislocation dipole configuration in the DFK model.  \protect\subref{fig:dipole_motion_comp_screw} compares the time-varying dislocation displacements predicted by atomistics and the DFK model. The slope of the plot yields the dislocation velocity. \protect\subref{fig:dipole_mobility_atom} compares the atomistic and the DFK model predictions of the variation of edge and screw dislocation velocities with applied shear traction. The slopes of the plots yield the dislocation drag coefficients.}
\label{fig:atom_mob_cal}
    \end{figure}

Beginning with an AB-stacked BG of size $L_x \times L_y = \SI{1476}{\angstrom} \times \SI{255.651}{\angstrom}$, a dislocation dipole of width $\approx \SI{170}{\angstrom}$ (see \fref{fig:dipole_schematic_atom}) is constructed by displacing atoms of the top lattice that lie in the region between the two dislocation lines by a Burgers vector $\bm b=(\SI{1.42}{\angstrom}) \bm e_2$. To avoid a singular dislocation core, the atoms are displaced according to a piecewise-linear displacement plotted in the bottom of \fref{fig:dipole_schematic_atom}. The continuum counterpart of the atomistic system is shown in \fref{fig:dipole_schematic_cont}. To construct the dislocation dipole in the DFK model, a deformation map $\bm \kappa_\alpha$ was designed according to the line plot shown in \fref{fig:dipole_schematic_atom} so that the dislocation core widths agree with those in atomistics. The imposed displacement field ensures the simulations consist of two AB regions separated by a central BA region. \fref{fig:dipole_schematic_cont} also shows the contribution of the elastic and vdW energies to the total energy. 

To obtain the inverse mobility of the continuum model, we first equilibrate the atomistic system and subsequently drive 
the dipole dislocations apart using MD by subjecting the atomistic system to shear traction of $\SI{3.296}{\mega\pascal}$ along the $X_2$ direction, as shown in \fref{fig:dipole_schematic_atom}. The plot in red in \fref{fig:dipole_motion_comp_screw} shows the evolution of a dislocation's position deduced by tracking atomic energies.  
Similarly, continuum dislocations are equilibrated and driven apart using a unit inverse mobility, and their positions are tracked using the energy density. Rescaling the inverse mobility to 
\begin{equation}
    b=\SI{1.215e-5}{\pascal\second\per\meter},
    \label{eqn:bfit}
\end{equation}
which is equivalent to rescaling the continuum timescale, results in identical evolution of the continuum (shown in black in \fref{fig:dipole_motion_comp_screw}) and atomistic dislocations. To ensure the inverse mobility in \eqref{eqn:bfit} is independent of system size, the above calculations were repeated for $L_x$ ranging from $\SI{1107}{\angstrom}$ to $\SI{1624}{\angstrom}$, resulting in inverse mobilities in the range 
$(1.215 \pm 0.038)\times 10^{-5}\si{\pascal\second\per\meter}$.

Next, we validate the fitted continuum model by repeating the above simulation for edge dislocations and varying shear tractions. \fref{fig:dipole_mobility_atom} compares the predictions of the atomistic and continuum models and shows excellent agreement. In particular, we note that a single choice of inverse mobility, given in \eqref{eqn:bfit}, successfully predicts the distinct kinetics of edge and screw dislocations. In addition, we record the dislocation drag $d$--- defined as the inverse of the slope of a plot in \fref{fig:dipole_mobility_atom} --- for the screw dislocation to be $\SI{0.0480}{\mega\pascal\second\per\meter}$. For comparison, this is $144$ times smaller than the screw dislocation drag in BCC tungsten at $\SI{0}{\kelvin}$  \citep{po2016phenomenological}.
Similarly, the edge dislocation drag coefficient is calculated as $\SI{0.0281}{\mega\pascal\second\per\meter}$. Therefore, at $0$ K temperature, an edge dislocation in BG is $1.7$ times more mobile than a screw dislocation. For comparison, the ratio of edge dislocation mobility to screw dislocation mobility is equal to $2.3$ and $2.0$ for BCC W and FCC Al, respectively \citep{po2016phenomenological,olmsted2005atomistic}.
\subsection{Computation of friction in heterodeformed BG using the DFK model}
\begin{figure}
    \centering
    \subfloat[Twisted BG]
    {
        \includegraphics[height=0.35\textwidth]{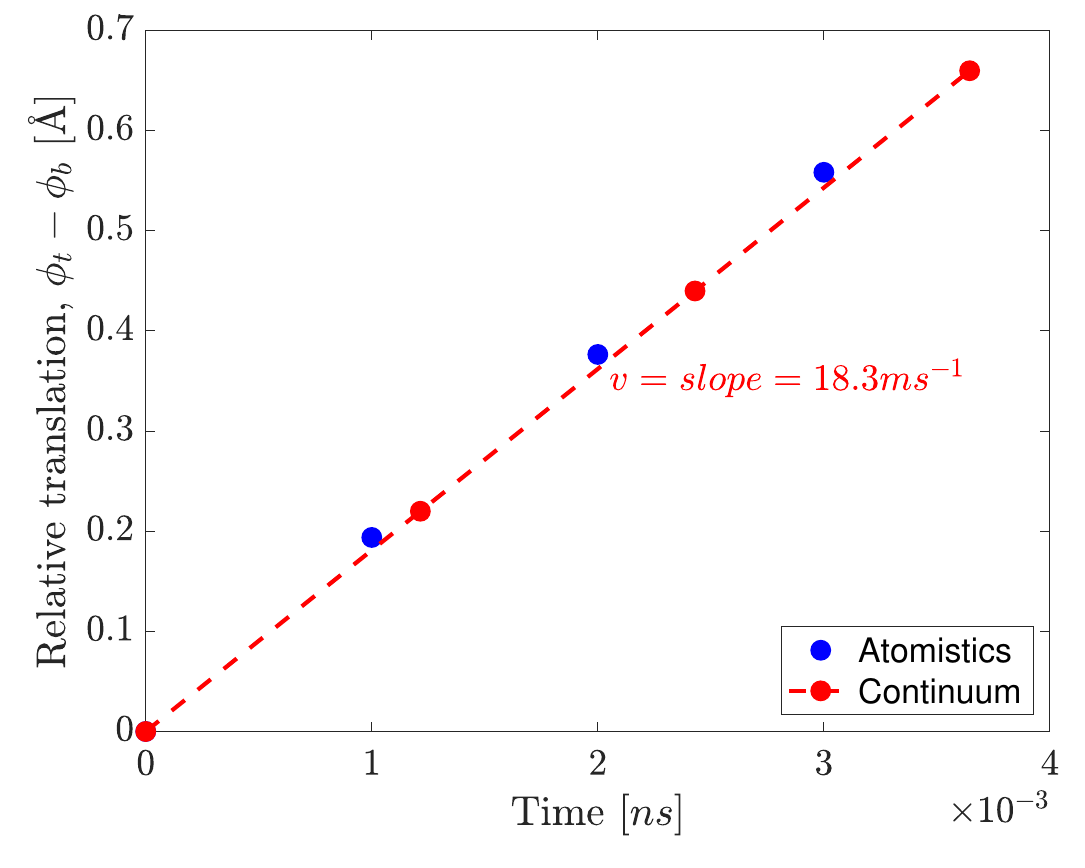}
        \label{fig:rel_tran_cont}
    }
    \subfloat[Equi-biaxial heterostrained BG]
    {
        \includegraphics[height=0.35\textwidth]{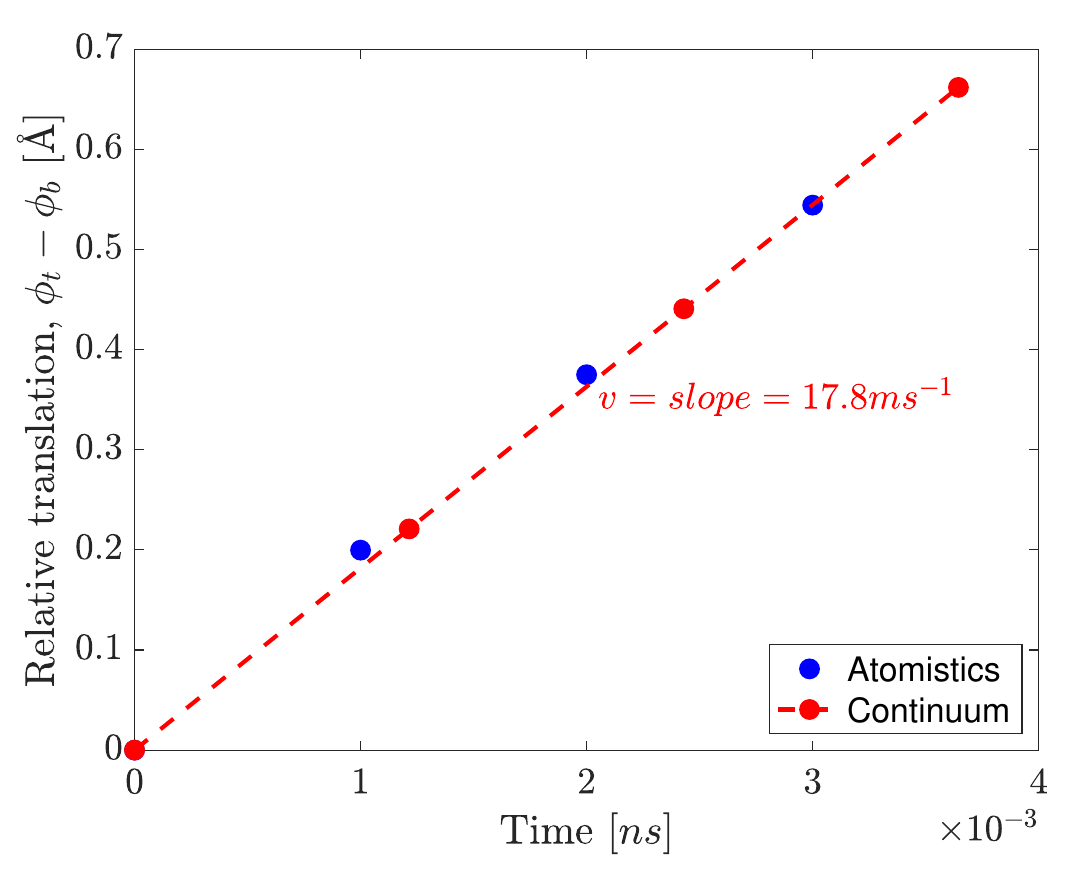}
        \label{fig:rel_tran_str_cont}
    }
    \\
    \subfloat[]
    {
        \includegraphics[height=0.35\linewidth]{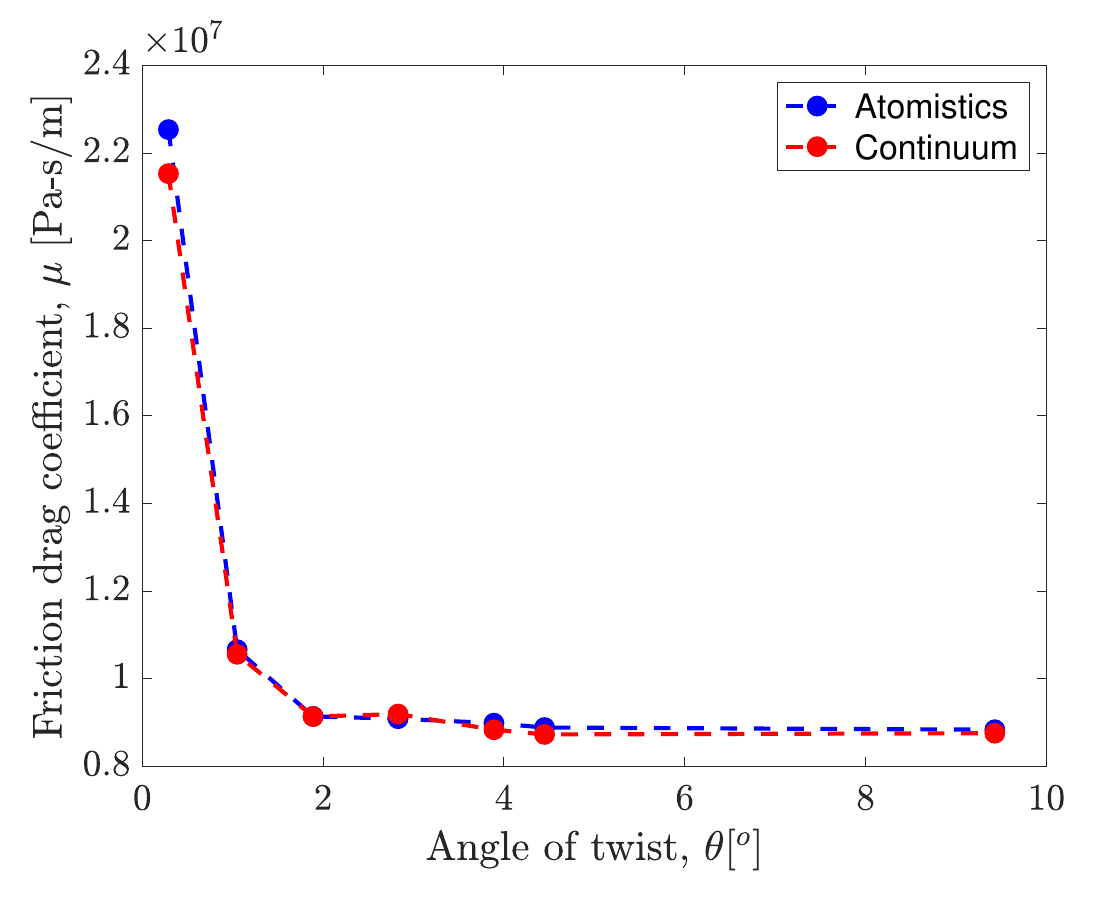}
        \label{fig:twist_atom_cont_friction}
    }
\caption{\protect\subref{fig:rel_tran_cont} and \protect\subref{fig:rel_tran_str_cont} show the time variation of relative translation between the layers in a $4.41^{\circ}$ twisted BG and a $1.8519\%$ equi-biaxial heterostrained BG, respectively, when subjected to a shear traction of $160$ MPa along the $X_2$ direction. \protect\subref{fig:twist_atom_cont_friction} compares the atomistic and continuum predictions of the variation of the friction drag coefficient with twist angle in twisted BGs.}
\label{fig:cont_result}
\end{figure}

Next, we use the fitted continuum model of the previous section to predict the drag coefficient in heterodeformed BGs.  \fref{fig:rel_tran_cont} compares the time evolutions recorded in the atomistic and continuum simulations of relative translation between the two layers of a $4.408455^{\circ}$ twisted BG subjected to shear traction of $\SI{160}{\mega\pascal}$ in the $\bm e_2$ direction. The relative sliding velocities --- inferred by the slope --- in the atomistic and continuum simulations are in agreement at $\SI{18.3}{\meter\per\second}$ and $\SI{18.0}{\meter\per\second}$, respectively. Results from a similar calculation for a $1.8519\%$ equi-biaxial heterostrained BG, plotted in \fref{fig:rel_tran_str_cont}, yielded relative translation velocities of $v= 17.8$ ms$^{-1}$ 
and $17.5$ ms$^{-1}$ in continuum and atomistic simulations, respectively. Repeating the above calculation for BGs with varying twist angles yields the friction drag as a function of the twist angle. The atomistic and continuum data plotted in \fref{fig:twist_atom_cont_friction} show excellent agreement confirming our assertion stated at the end of \sref{sec:friction} that friction drag can be predicted from the mobility of dislocations. 

\section{Summary and Conclusions}
\label{sec:conclusion}
In this paper, we developed an atomistically informed dynamic Frenkel--Kontorova (DFK) continuum model to predict the friction drag coefficient in heterodeformed  2D bilayer systems. The DFK model was motivated by atomistic simulations of heterodeformed bilayer graphene (BG) that suggested friction emerges from the intrinsic drag of interface dislocations that form during structural relaxation. In particular, MD simulations of twisted and equi-biaxial heterostrained BGs subjected to a shear force show that the network of dislocations translates in unison with a constant velocity. Using the DFK model, we demonstrated that the friction drag coefficient is a property that emerges from the geometry of the network of interface dislocations and their intrinsic drag. 
The key features of the DFK model are
\begin{enumerate}
    \item The unknown primitive variables of the model are the displacement fields of the two continuum layers of a bilayer. Dislocations are diffused and derived features originating from jumps (large gradients) in the displacement fields.
    \item Dissipation is governed by a single inverse mobility constant, $b$, associated with the rate of change of the displacement fields. 
    \item In addition to the elastic stiffness of the layers, the inputs to the model are the generalized stacking fault energy of the AB stacking and the inverse mobility $b$. The latter was fit using an atomistic simulation of an expanding screw dislocation dipole.
    \item A single choice of $b$ successfully predicts interfacial friction for \emph{any} heterodeformation, thereby overcoming the limitation of the conventional Prandtl--Tomlinson (PT) model fit to a specific heterodeformation.
\end{enumerate}
While recent developments in constructing interatomic potentials make atomistic simulations reliable for predicting twist-dependent friction drag, the large four-dimensional heterodeformation space presents a significant computational challenge. Furthermore, because of the time constraints associated with atomistic simulations, their application is limited to investigating sliding velocities much larger than those encountered in practical scenarios. Therefore, using the DFK model to map friction drag as a function of the four-dimensional heterodeformation space renders it a high-throughput tool for a large-scale investigation into the frictional properties of heterodeformed BG interfaces.

Before concluding, we note the limitations of the current study. First, the DFK model was calibrated for high sliding velocities ($>\SI{1e-3}{\meter\per\second}$), where the friction drag coefficient is constant. However, the coefficient varies logarithmically \citep{wang2024colloquium} at lower velocities with the applied load as the dislocation motion is thermally activated \citep{proville2020modeling}. Extending the application of the DFK model to the low-velocity regime is straightforward and requires incorporating an inverse mobility dependent on the shear-traction and temperature. Second, DFK model does not incorporate out-of-plane displacement. 
Consequently, the size of the dislocation junction/AA stacking predicted by the DFK model is slightly larger than that observed in atomistic simulations. 
Moreover, out-of-plane displacement can transform the straight dislocation lines into helical lines, signifying its notable influence on the dislocation network \citep{dai2016twisted,rakib2023helical}. To incorporate the out-of-plane displacement, the constitutive law of the DFK model should include a) a 3D GSFE \citep{zhouVanWaalsBilayer2015} (as opposed to the current 2D GSFE), wherein the third dimension corresponds to the interlayer spacing, and b) bending \citep{dai2016structure} rigidity of the constituent 2D materials. Finally, we restricted our study to twisted and equi-biaxial heterostrained BGs, wherein the friction is antiparallel to the relative velocity, resulting in a scalar drag coefficient. However, more general heterodeformations may yield drag coefficients with a tensorial character. In future work, we will apply the DFK model to map the tensorial drag coefficient over the entire heterodeformation space.
\begin{acknowledgement}

NCA and TA would like to acknowledge support from the National Science Foundation Grant NSF-MOMS-2239734 with S. Qidwai as the program manager. This research used the Delta advanced computing and data resource which is supported by the National Science Foundation (award OAC 2005572) and the State of Illinois. Delta is a joint effort of the University of Illinois Urbana-Champaign and its National Center for Supercomputing Applications.

\end{acknowledgement}

\begin{suppinfo}
The following files are available free of charge.
\begin{itemize}
  \item SuppInformation.pdf: Details on the implementation of continuum substrate in LAMMPS, and the implementation of the dynamic Frenkel--Kontorova model.
  \item twistBGshearing.mp4: Motion of the network of screw dislocations in a $1.050121^\circ$ twisted BG subjected to shear traction along the $X_2$ direction.
  \item heterostrainBGshearing.mp4: Motion of the network of interface dislocations in a $0.4219\%$ equi-biaxial heterostrained BG under a shear traction along the $X_1$ direction.
\end{itemize}

\end{suppinfo}

\bibliography{achemso-demo}

\end{document}